\documentclass[a4paper,11pt]{article}
\pdfoutput=1
\usepackage{jheppub}
\usepackage[T1]{fontenc} 
\usepackage{fixltx2e}

\usepackage{pdflscape}
\usepackage{afterpage}
\usepackage{vmargin,amssymb,amsmath,epsfig,latexsym,booktabs,braket}

\newcommand{\cN}{{\cal N}}

\newcommand{\cO}{{\cal O}}

\let\emptyset\varnothing

\definecolor{darkgreen}{rgb}{0,0.5,0}

\newcommand{\one}{\textcolor{blue}{1}\,}
\newcommand{\two}{\textcolor{red}{2}\,}
\newcommand{\three}{\textcolor{darkgreen}{3}\,}

\title{\boldmath Positivity of hexagon perturbation theory}

\preprint{{\small HU-EP-18/18, HU-MATH~2018-05,
TCDMATH~18-08}}

\author[a]{Burkhard Eden,}
\author[b]{Yunfeng Jiang,}
\author[c]{Marius de Leeuw,}
\author[a]{Tim Meier,}
\author[a]{Dennis le Plat,}
\author[b]{Alessandro Sfondrini}

\affiliation[a]{Institut f\"{u}r Mathematik und Physik, Humboldt-Universit{\"a}t zu Berlin, \\
Zum gro{\ss}en Windkanal 6, 12489 Berlin, Germany}
\affiliation[b]{Institut f\"ur theoretische Physik, ETH Z\"urich, \\
Wolfgang-Pauli-Stra{\ss}e 27, 8093 Z\"urich, Switzerland}
\affiliation[c]{School of Mathematics
Trinity College Dublin
Dublin, Ireland}

\emailAdd{eden@math.hu-berlin.de}
\emailAdd{jiangyu@phys.ethz.ch}
\emailAdd{mdeleeuw@maths.tcd.ie}
\emailAdd{tmeier@physik.hu-berlin.de}
\emailAdd{diplat@physik.hu-berlin.de}
\emailAdd{sfondria@itp.phys.ethz.ch}

\abstract{
The hexagon-form-factor program was proposed as a way to compute three- and higher-point correlation functions in $\mathcal{N}=4$ super-symmetric Yang-Mills theory and in the dual AdS$_5\times$S$^5$ superstring theory, by exploiting the integrability of the theory in the 't~Hooft limit. This approach is reminiscent of the asymptotic Bethe ansatz in that it applies to a large-volume expansion. Finite-volume corrections can be incorporated through  L\"uscher-like formulae, though the systematics of this expansion is largely unexplored so~far. Strikingly, finite-volume corrections may feature negative powers of the 't~Hooft coupling~$g$ in the small-$g$ expansion, potentially leading to a breakdown of the formalism.
In this work we show that the finite-volume perturbation theory for the hexagon is positive and thereby compatible with the weak-coupling expansion for arbitrary $n$-point functions.}

\begin{document}
\maketitle
\flushbottom

\section{Introduction}
\label{sec:intro}
Computing correlation functions is one of the central problems in quantum field theory. For a generic interacting theory, it is impossible to calculate such observables non-perturbatively. Usually one needs to perform an expansion in certain parameters. A typical example is the small-coupling expansion around a free theory, which can be performed by well-established techniques such as Feynman diagrams.
In some special theories, more powerful alternative techniques may exist which are usually based on symmetries.  In integrable quantum field theories, the form-factor bootstrap approach is such an alternative. %
%
%
%
 While most integrable models appear in one or two dimensions, integrability sometimes manifests itself in higher-dimensional theories too, for instance through dualities such as AdS/CFT~\cite{Maldacena:1997re, Witten:1998qj,Gubser:1998bc}. One of the most prominent examples is the $\mathcal{N}=4$ supersymmetric Yang-Mills theory (SYM), dual to type-IIB superstrings on AdS$_5\times$S$^5$. In the planar limit~\cite{'tHooft:1973alw}, the integrability of the spectral problem for the two-dimensional theory on the string worldsheet carries over to $\mathcal{N}=4$ SYM in the guise of an integrable spin~chain~\cite{Minahan:2002ve}, see \textit{e.g.}\  refs.~\cite{Arutyunov:2009ga,Beisert:2010jr} for reviews.
Interestingly, it was recently realised that integrability might play a role for more general observables too.
 Indeed, a generalisation of the form factor approach to $\mathcal{N}=4$ SYM has been proposed in ref.~\cite{Basso:2015zoa} in terms of hexagonal tessellations; in the following we refer to this construction as the \emph{hexagon approach}. This was initially proposed as a way to compute planar three-point functions involving non-protected operators. Soon after, it was realised that those techniques could be adapted to higher-point planar correlation functions~\cite{Eden:2016xvg, Fleury:2016ykk} and at least to some extent to non-planar observables~\cite{Eden:2017ozn, Bargheer:2017nne}.%
\footnote{Non-planar observables in $\cN =4$ SYM are also being actively investigated by other integrability-based approaches, see \textit{e.g.}\ refs.~\cite{Carlson:2011hy,deMelloKoch:2018tlb}, as well as  ref.~\cite{Kristjansen:2010kg} for a review of earlier developments.}

The hexagon approach is reminiscent of the ``asymptotic'' Bethe ansatz for the $\mathcal{N}=4$ SYM spin~chain~\cite{Beisert:2005fw}, or of the Bethe-Yang equations in two dimensional integrable QFTs: it is only exact up to exponentially-small corrections in the volume of the theory --- the $R$-charge of the $\mathcal{N}=4$ SYM operators under consideration. In the spectral problem, such finite-size corrections can be interpreted as due to virtual (``mirror'') particles wrapping the worlsdheet~\cite{Ambjorn:2005wa}, similar to the L\"uscher effects of relativistic theories~\cite{Luscher:1985dn,Luscher:1986pf}. In the hexagon program, they can similarly be described as mirror particles probing the finite-size structure of the hexagon tessellation. For three-point functions, the first few finite-volume corrections can be explicitly computed and matched with small-coupling perturbation theory~\cite{Basso:2015zoa,Eden:2015ija, Basso:2015eqa,Basso:2017muf}.
Furthermore, in that context only a finite number of finite-volume corrections needs to be computed at each given order of the 't~Hooft coupling expansion~\cite{Basso:2015eqa}.
The situation is less clear for higher-point functions. The results of the hexagon approach for four-~\cite{Eden:2016xvg, Fleury:2016ykk} and five-point~\cite{Fleury:2017eph} functions and for certain non-planar observables~\cite{Eden:2017ozn, Bargheer:2017nne} match with the lowest orders of perturbation theory; such integrability computations account for the first few finite-volume corrections. Still, it was not shown so far that more complicated finite-volume effects (involving more mirror particles) can indeed be neglected at those orders.
More generally, we do not know which finite-volume effects (\textit{i.e.}\ how many mirror particles) we need to take into account at a given order of the 't~Hooft-coupling expansion. Establishing this relation for general correlation functions is the aim of this paper.

The reason why this is a subtle problem --- compared for instance to L\"uscher corrections in the spectral problem --- is that in the hexagon formalism mirror particle come in \textit{three} distinct families. Indeed three of the six edges of the hexagon correspond to physical excitations, and the remaining three correspond to mirror (or anti-mirror) ones, leading to distinct kinematic regions. In the hexagon form factor, while physical processes yield non-negative powers of the 't~Hooft coupling~$g^2$ in the $g\ll1$ expansion, mirror-anti-mirror processes yield a factor of~$g^{-2}$. Na\"ively, populating a hexagon with sufficiently many mirror magnons can make a process which is extremely suppressed in ``large-volume'' appear \textit{even at tree level}~$g^0$. Even worse, in principle we might expect contributions to processes \textit{with arbitrarily negative powers of $g^2$!}

Clearly such a situation would be disastrous for the hexagon program. Since several explicit computations match field theory results~\cite{Eden:2016xvg, Fleury:2016ykk, Fleury:2017eph, Eden:2017ozn, Bargheer:2017nne}, there should be a way to resolve this apparent issue. One possibility is that such multi-mirror-magnon processes indeed appear, but somehow cancel at appropriate orders in $g^2$  for correlation functions of physical states; this would make the formalism consistent in principle, but it would also make it almost impossible to extract physical data from the hexagon construction for generic observables without knowing \textit{a priori} which processes cancel and which do~not. The other much more appealing possibility is that there exist a refined estimate where, as we add more and more mirror magnons, we obtain higher and higher terms in the~$g^2$ expansion. In this paper we show that this is the case \textit{for arbitrary (planar and non-planar) correlation functions} of protected operators, obtaining an explicit formula relating the number of mirror magnons to the order in the small-$g$ expansion. We call this property the \textit{positivity} of the hexagon perturbation theory. This also extends to the case of non-protected operators, which however depends more subtly on which particular correlator we consider.

The paper is structured as follows:
we start by reviewing some essential features of the hexagon approach in section~\ref{sec:review_hexagon}. In section~\ref{sec:positivity1} we present our main result: an improved estimate for the contribution of mirror processes to the $g$-expansion, which is \textit{bounded from below}, valid for any correlator of BPS operators; we also comment on the extension to non-BPS operators. In section~\ref{sec:comments} we apply these ideas to the computation of planar four-point functions of BPS operators. We conclude in section~\ref{sec:conclusions}. We also present an alternative (algorithmic) derivation of our improved bound in appendix~\ref{sec:positivity2}.

\section{Review of hexagon form factors}
\label{sec:review_hexagon}
When computing four-point and higher correlation functions we need to triangulate a punctured Riemann surface (the sphere, for planar correlators). The edges of such triangles are the ``mirror'' edges of the hexagons. By blowing up the punctures to small circles we can add three more ``physical'' edges corresponding to arcs on those circles. Since we are interested in obtaining an estimate valid \textit{for any correlation functions}, below we shall not make any assumption on how hexagons are glued together. Instead, we will consider a single hexagon.

\subsection{Form~factor for a single hexagon}

As we mentioned, the key idea of the hexagon approach to correlation functions is to decompose any correlation function into a hexagonal tessellation. To ``glue'' together such a tessellation it is necessary to insert a complete basis of states for each of the three ``mirror'' edges. This amounts to populating the hexagon with mirror magnons. Notice that, precisely because mirror magnons arise from ``gluing'', they are shared by two contiguous hexagons. Taking this into account, the hexagon form factor with $\mathbf{u},\mathbf{v},\mathbf{w}$ sets of mirror magnons and $\mathbf{x},\mathbf{y},\mathbf{z}$ sets of physical magnons takes the form
\begin{align}
\label{eq:hexagonschematic}
\mathrm{hexagon} \sim
\sqrt{\mu_{\mathbf{u}}\mu_{\mathbf{v}}\mu_{\mathbf{w}}}\,e^{\frac{i}{2} \tilde{E}_{\mathbf{u}}\ell_{\mathbf{u}}+\dots}\,e^{i \psi \mathbf{J}+\dots}\,
\langle \,\mathfrak{h}\,|\,\mathbf{z}^{4\gamma},\mathbf{u}^{3\gamma},\mathbf{y}^{2\gamma},\mathbf{v}^{\gamma},\mathbf{x}^{0\gamma},\mathbf{w}^{-\gamma}\rangle\,.
\end{align}
Some comments are in order. Here $\mu$ is the mirror measure corresponding to each mirror magnon; we only assign ``half'' of such a measure to each hexagon, as mirror magnons are shared across two hexagons. For the same reason, we split in half the ``bridge length'' $\ell$ along which each magnons propagates. Next, we allow for an arbitrary chemical potential $\psi$ (for any possible magnon charge) which is needed to describe four- and higher-point functions~\cite{Eden:2016xvg,Fleury:2016ykk}.
The powers of $\gamma$ identify how many time we perform the mirror transformation in the notation of ref.~\cite{Basso:2015eqa}. Notice that physical particles sit at even-$\gamma$ edges, and mirror ones at odd-$\gamma$ ones.
Finally,%
\footnote{In order to compute a complete correlation function it would be necessary to consider all hexagons in the tessellation, sum over partitions for physical excitations, sum over all possible mirror particles and integrate over their rapidities --- and normalise the result appropriately. The detail of this, as we mentioned, depend on the particular correlator we are computing; here we will not perform this procedure, as we aim at an estimate as general as possible.
}
 the hexagon form factor~$\mathfrak{h}$ can then be related to the centrally extended $\mathfrak{su}(2|2)$ (bound state) $S$-matrix $S_{ij}$ \cite{Beisert:2005tm,Arutyunov:2009mi}
\begin{align}
\langle\, \mathfrak{h}\,|\,\mathbf{z}^{4\gamma},\mathbf{u}^{3\gamma},\mathbf{y}^{2\gamma},\mathbf{v}^{\gamma},\mathbf{x}^{0\gamma},\mathbf{w}^{-\gamma}\rangle
 \sim  \prod_{i<j} h_{ij}  S_{ij}\, ,
\end{align}
up to a suitable projection in flavour space on the right-hand side~\cite{Basso:2015zoa}.
The hexagon factor $h_{ij}$ for two bound states with bound state numbers $Q_1,Q_2$ is given by
\begin{align}
h_{12}&= \prod_{k=-\frac{Q_1-1}{2}}^{\frac{Q_1-1}{2}}\prod_{l=-\frac{Q_2-1}{2}}^{\frac{Q_2-1}{2}}h(u_1^{[2k]},u_2^{[2l]})\,,\\
h(u_1,u_2) &= \frac{x_1^--x_2^-}{x_1^--x_2^+}\frac{1-\frac{1}{x^-_1x^+_2}}{1-\frac{1}{x^+_1x^+_2}}\frac{1}{\sigma(u_1,u_2)}\,.
\end{align}
The dressing factor $\sigma_{12}$ is given by the BES phase \cite{Beisert:2006ez}. The purpose of this paper is to explore whether a perturbative weak-coupling expansion is compatible with the hexagon approach. In order to address this question we need to understand the weak-coupling expansion of
\begin{equation}
\mathcal{S}_{ij} =  h_{ij} S_{ij}\,,
\end{equation}
\textit{i.e.}\ of the $su(2|2)$ $S$-matrix dressed by the hexagon scalar factor, in different kinematic channels.

\subsection{Weak-coupling expansions}
In order to see how different pieces of eq.~\eqref{eq:hexagonschematic} scale when $g\ll1$, let us collect here the weak-coupling expansion of the relevant quantities, and fix our conventions.

\paragraph{Conventions,} We need to compute the $S$-matrix between magnons with different mirror orientations. For consistency, we will work with the \textit{string-frame} $S$-matrix~\cite{Arutyunov:2006yd}, see also ref.~\cite{Arutyunov:2009ga} for a review.
We follow the conventions of \cite{Basso:2015zoa,Fleury:2017eph}, where the bound-state $S$-matrix is normalized such that the scattering of the highest-weight fermionic state is set to unity. Moreover, we note that the dressing factor $\sigma$ for two bound states with bound-state numbers $Q_{1,2}$ satisfies the following crossing equations \cite{Arutyunov:2009kf}
\begin{align}
\sigma_{12}(u_1^{2^\gamma},u_2) &= \Big(\frac{x_2^-}{x_2^+}\Big)^{Q_1} \frac{x_1^--x^+_2}{x_1^--x^-_2}\frac{1-\frac{1}{x^+_1x^+_2}}{1-\frac{1}{x^+_1x^-_2}}
\prod_{k=1}^{Q_1-1}\frac{u_1-u_2-i\frac{Q_2-Q_1+2k}{g}}{u_1-u_2+i\frac{Q_2-Q_1+2k}{g}} \frac{1}{\sigma_{12}(u_1,u_2) },\\
\sigma_{12}(u_1,u_2^{-2^\gamma}) &= \Big(\frac{x_1^+}{x_1^-}\Big)^{Q_2} \frac{x_1^--x^+_2}{x_1^--x^-_2}\frac{1-\frac{1}{x^+_1x^+_2}}{1-\frac{1}{x^+_1x^-_2}}
\prod_{k=1}^{Q_1-1}\frac{u_1-u_2-i\frac{Q_2-Q_1+2k}{g}}{u_1-u_2+i\frac{Q_2-Q_1+2k}{g}} \frac{1}{\sigma_{12}(u_1,u_2) }.
\end{align}
The dressing factor respects unitarity so that $\sigma_{12} = 1/\sigma_{21}$. By using the crossing equations and unitarity we can, for instance, relate all the dressing phases between mirror and anti-mirror particles to $\sigma(u^\gamma,v^\gamma)$.

\paragraph{Expansions.} Let us spell out the expansions of the different scattering matrices $\mathcal{S}$ for small coupling $g$.
These can be straightforwardly checked for the scattering of fundamental magnons and, by the fusion procedure, the results carry over to bound-state $S$-matrices.%
\footnote{%
We have explicitly checked all these expansions for the $Q=2$ bound-state $S$-matrix.
}
For virtual particles with the same mirror orientation, we find that the expansion starts at order $g^2$
\begin{align}
\mathcal{S}(u_i^{3\gamma},u_j^{3\gamma}) \sim \mathcal{S}(u_i^{\gamma},u_j^{\gamma}) \sim \mathcal{S}(u_i^{-\gamma},u_j^{-\gamma}) \sim g^2.
\end{align}
For physical particles, we find that the expansion always starts at order $1$
\begin{align}
&\mathcal{S}(u_i,u_j) \sim \mathcal{S}(u_i^{2\gamma},u_j^{2\gamma}) \sim \mathcal{S}(u_i^{4\gamma},u_j^{4\gamma}) \sim g^0\\
&\mathcal{S}(u^{2\gamma}_i,u_j) \sim \mathcal{S}(u^{4\gamma}_i,u_j) \sim \mathcal{S}(u_i^{4\gamma},u_j^{2\gamma}) \sim g^0.
\end{align}
Next we expand the $S$-matrix for the scattering processes that involve virtual magnons from different edges
\begin{align}
\mathcal{S}(u_i^{3\gamma},v_j^{\gamma}) \sim \mathcal{S}(u_i^{3\gamma},u_j^{-\gamma}) \sim \mathcal{S}(u_i^{\gamma},u_j^{-\gamma})  \sim g^{-2}.
\end{align}
This is the contribution that makes it possible, at least in a na\"ive estimate, to obtain arbitrarily negative powers of~$g^2$ when adding mirror magnons.
Finally, let us also consider the scattering between a virtual and a physical magnon. The dressed $S$-matrix is of order~1 unless the virtual magnon is on the edges that are opposite to the physical magnon%
\begin{align}\label{eq:phys-virtual}
\mathcal{S}(u^{n\gamma}_i,u_j^{(n+3)\gamma}) \sim
 \left\{\begin{array}{ll} g^{-1} & \text{if magnon 1 is a boson $\phi^a$,}\\g^{0} &\mathrm{if~ magnon~1~ is~a ~fermion~\psi^\alpha},\end{array}\right.
\end{align}
where $n=-2,0,2$.%
\footnote{%
More precisely, processes where fermions are transmitted are always order~1; reflection and boson-fermion processes are generically of order $g^{-1/2}$, and processes with physical bosons give more negative powers of~$g$. The precise scaling depends on choice of the magnon basis, in particular on the fermion-normalisation~$e^{ip/4}\sqrt{i(x^+_p-x^-_p)}$~\cite{Arutyunov:2006yd}. In the conventional basis bosons scale worst, like $g^{-1}$, while there exists a basis where all physical-fermion processes are of order~1. Final results are basis-independent.
}
 In the last two expansions we encounter yet another potential problem, as the negative powers of the coupling constant $g$ can potentially lead to arbitrary negative powers of the coupling constant in the hexagon expansion.

\paragraph{Multiple scattering.} A remarkable observation which will be crucial in what follows is that the weak-coupling expansion of the scattering process between three virtual magnons may scale ``better than na\"ively expected''. Specifically, let us take three mirror magnons from three different edges. We then find
\begin{align}
\label{eq:triangle}
\mathcal{S}_{12}(u^{3\gamma},v^{+\gamma})\, \mathcal{S}_{13}(u^{3\gamma},w^{-\gamma})\,\mathcal{S}_{23}(v^{+\gamma},w^{-\gamma})\, \sim\, g^{-2}.
\end{align}
We would expect this process to scale like $g^{-6}$, but it turns out that the most divergent contributions cancel out, so that the process scales \textit{four orders higher} in the~$g$ expansion than we would expect!
This is a rather striking and unique property of the $S$-matrix: we have verified that, up to scattering six magnons, there are no further identities of this type in any mirror channel.
Finally, it is also interesting to note that a similar (weaker) triple-scattering property arises also in the case where one of the magnons is physical, one virtual magnon is adjacent and the other virtual magnon is across the physical magnon,
\begin{equation}
\begin{aligned}
\mathcal{S}_{12}(u^{3\gamma},v^{+\gamma})\, \mathcal{S}_{13}(u^{3\gamma},x^{0\gamma})\,\mathcal{S}_{23}(v^{+\gamma},x^{0\gamma})\ &\sim& g^{-2},\\
 \mathcal{S}_{12}(u^{3\gamma},x^{0\gamma})\,\mathcal{S}_{13}(u^{3\gamma},w^{-\gamma})\,\mathcal{S}_{23}(x^{0\gamma},w^{-\gamma})\ &\sim& g^{-2}.
\end{aligned}
\end{equation}
According to \eqref{eq:phys-virtual} this could have been of order $g^{-3}$ when the physical magnon is a boson. What is more, the above relation can be generalized to an arbitrary number of physical magnons on the $0\gamma$ edge:
\begin{equation}
\label{eq:trianglephys}
\begin{aligned}
\mathcal{S}_{12}(u^{3\gamma},v^{+\gamma})
\Big[\prod_{j}^{\rightarrow} \mathcal{S}_{1j}(u^{3\gamma},x_j^{0\gamma})\Big]\Big[\prod_j^{\rightarrow}\mathcal{S}_{2j}(v^{+\gamma},x_j^{0\gamma})\Big]\ &\sim& g^{-2},\\
\Big[\prod_{j}^{\rightarrow}  \mathcal{S}_{1j}(u^{3\gamma},x_j^{0\gamma})\Big]\, \mathcal{S}_{12}(u^{3\gamma},w^{-\gamma}) \Big[\prod_{j}^{\leftarrow} \mathcal{S}_{j2}(x_j^{0\gamma},w^{-\gamma})\Big]\ &\sim& g^{-2},
\end{aligned}
\end{equation}
where $\displaystyle\prod^{\rightarrow} \mathcal{S}_{aj}= \mathcal{S}_{a1} \mathcal{S}_{a2}\ldots$ and $\displaystyle\prod^{\leftarrow} \mathcal{S}_{aj}= \mathcal{S}_{aN} \mathcal{S}_{a,N-1}\ldots$ . Similar expressions hold for physical magnons on the $2\gamma$ and $4\gamma$~edges, or indeed combining physical magnons of different types:
\begin{equation}
\begin{aligned}
\label{eq216}
&\Big[\prod_{i}^\leftarrow \mathcal{S}_{1i}(u^{3\gamma},y_i^{2\gamma})\Big]\!
\Big[\prod_{j}^\leftarrow \mathcal{S}_{1j}(u^{3\gamma},x_j^{0\gamma})\Big]
\mathcal{S}_{12}(u^{3\gamma},w^{-\gamma})\times \\
&\qquad \qquad \qquad \qquad\times\Big[\prod_j^\rightarrow \mathcal{S}_{j2}(x_j^{0\gamma},w^{-\gamma})\Big]\!
\Big[\prod_i^\rightarrow \mathcal{S}_{i2}(y_i^{2\gamma},w^{-\gamma})\Big] \sim\, g^{-2}
\end{aligned}
\end{equation}
%

\section{Positivity of hexagon perturbation theory}
\label{sec:positivity1}
Let us now derive an explicit formula for the order in~$g\ll1$ at which a given hexagon configuration starts to contribute to the perturbative expansion. We first consider the case where we have only mirror excitations on the three mirror edges; this is the relevant case when computing arbitrary correlation functions of half-BPS operators.

\subsection{Na\"ive estimate of of 't~Hooft-coupling order}
The contribution of a single hexagon~\eqref{eq:hexagonschematic} is a little simpler for BPS operators
\begin{align}
\label{eq:hexagonschematicmirror}
\sqrt{\mu_{\mathbf{u}}\mu_{\mathbf{v}}\mu_{\mathbf{w}}}\,e^{\frac{i}{2} \tilde{E}_{\mathbf{u}}\ell_{\mathbf{u}}+\dots}\,e^{i \psi \mathbf{J}+\dots}\,
\langle\, \mathfrak{h}\,|\emptyset,\mathbf{u}^{3\gamma},\emptyset,\mathbf{v}^{\gamma},\emptyset,\mathbf{w}^{-\gamma}\rangle = O(g^{p})\,.
\end{align}
This scales like $g^{p}$ when $g\ll1$,  where the order $p$ is, na\"ively
\begin{equation}
\label{eq:pnaive}
p_{\text{na\"ive}}=\sum_{i=1}^3N_i(1+\ell_i)+2\sum_{i=1}^3\frac{N_i(N_i-1)}{2}-2N_1N_2-2N_1 N_3-2N_2 N_3\,,
\end{equation}
where we have introduced the short-hand notations
\begin{align}
|\mathbf{u}|=N_1,\qquad |\mathbf{v}|=N_2,\qquad |\mathbf{w}|=N_3.
\end{align}
Let us see how the different terms arise. Firstly, the first term in eq.~\eqref{eq:pnaive} is given by the contribution of ``half the measure'' which scales as~$g^1$ for each mirror magnon, plus half of the contribution of the bridge lengths. Next, when scattering all particles on the $i$th mirror edge among themselves, we have a total of~$N_i(N_i-1)/2$ processes, each contributing~$g^2$. Finally, and \textit{herein lies the rub}, when scattering magnons from different edges we get \textit{negative} contributions, of order $g^{-2}$ for each scattering event. We can rewrite the na\"ive estimate~\eqref{eq:pnaive} as
\begin{equation}
\begin{aligned}
p_{\text{na\"ive}}&=&\sum_{i=1}^3N_i\ell_i+\sum_{i=1}^3N_i^2-2N_1N_2-2N_1 N_3-2N_2 N_3\\
&=&\sum_{i=1}^3N_i\ell_i+(N_1-N_2)^2+N_3^2-2N_3(N_1+N_2)\,,
\end{aligned}
\end{equation}
which is clearly \textit{unbounded from below}, for instance when $N_1\sim N_2\gg N_3\gg \ell_i$.

\paragraph{Improving the estimate.}
Bearing in mind the observation of eq.~\eqref{eq:triangle}, we can get a better estimate for the scaling of a single hexagon, namely
\begin{equation}
\label{eq:pcomplete}
p=p_{\text{na\"ive}}+4T(N_1,N_2,N_3)\,,
\end{equation}
where $T(N_1,N_2,N_3)$ is the number of ``triple'' scattering events such as the ones of eq.~\eqref{eq:triangle}. We are therefore interested in arranging the scattering in such a way as to maximise such triple scattering processes.

\begin{figure}[t]
\begin{center}
\includegraphics[scale=0.22]{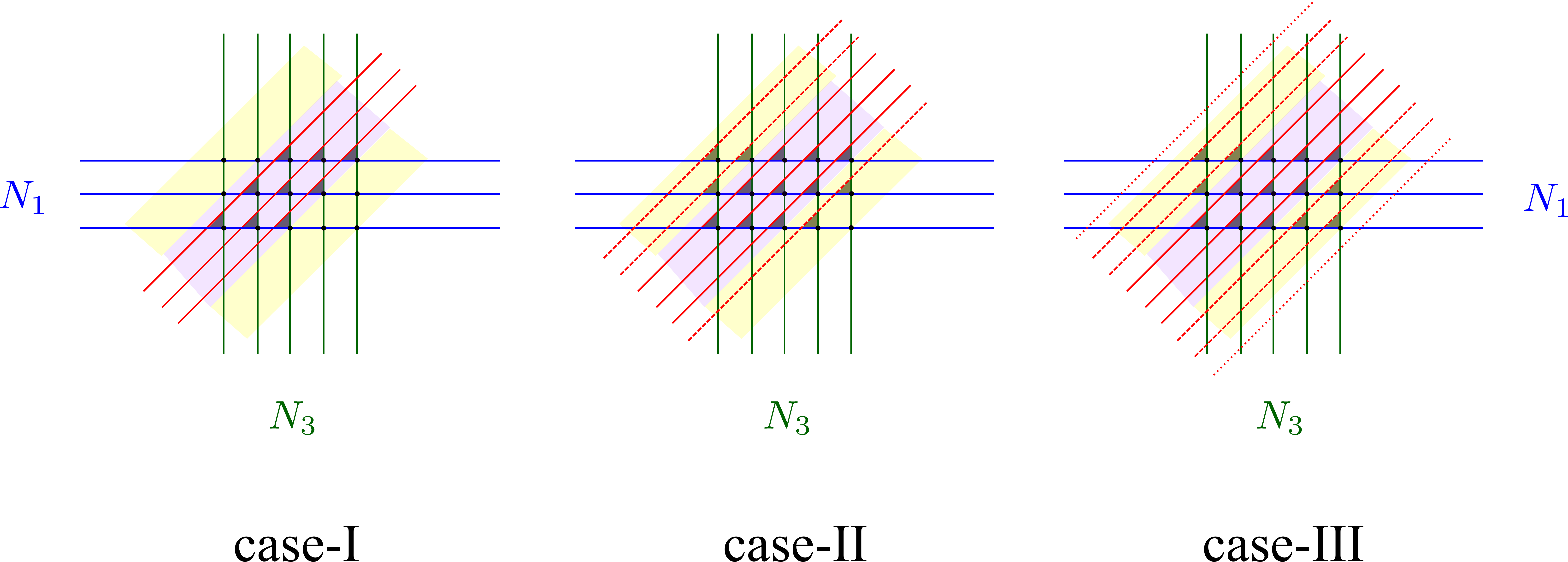}
\caption{Diagrammatic representation of scattering processes involving three sets of mirror magnons labelled by $N_1$, $N_2$ and $N_3$, corresponding to particles of type $\mathbf{u}$, $\mathbf{v}$ and $\mathbf{w}$, respectively.
We start by drawing the lines corresponding to $N_1$ and $N_3$ (assuming for definiteness $N_1\leq N_3$) and then place, one by one, the diagonal lines corresponding to $N_2$ in such a way as to maximise the triple scattering of eq.~\eqref{eq:triangle} --- the dark triangles.
As detailed in the main text, we see three distinct behaviours depending on how many diagonal lines we need to place: for the first few, which we place in the purple region (see left panel), we get a constant number of triangles per diagonal line. After a certain point (yellow region, middle panel), the number of triangles decreases as we add lines. Eventually, adding more diagonal lines does not yield any new triangle (right panel).}
\label{fig:triangle_diagram}
\end{center}
\end{figure}

\subsection{Triple scattering processes}
It is convenient to introduce a diagrammatic representation of the scattering processes which we will consider. In practice, we have three sets of lines $\{\mathbf{u}\}$,  $\{\mathbf{v}\}$,  $\{\mathbf{w}\}$, which we want to rearrange by reversing their order,
\begin{equation}
(u_1,\dots u_{N_1},v_1,\dots v_{N_2},w_1,\dots w_{N_3})\to
(w_{N_3},\dots w_1,v_{N_2},\dots v_{1},u_{N_1},\dots u_1).
\end{equation}
For the purpose of counting the triple-intersection of lines of type $u, v$ and $w$ the rearrangement within each set  $\{\mathbf{u}\}$,  $\{\mathbf{v}\}$,  $\{\mathbf{w}\}$ is irrelevant. Hence we can consider the slightly simpler process
\begin{equation}
(u_1,\dots u_{N_1},v_1,\dots v_{N_2},w_1,\dots w_{N_3})\to
(w_{1},\dots w_{N_3},v_{1},\dots v_{N_2},u_{1},\dots u_{N_1}).
\end{equation}
We have represented such a scattering process in figure~\ref{fig:triangle_diagram}. We start by drawing $N_1$ horizontal and $N_3$ vertical lines, which we can do unambiguously. For definiteness, we shall assume
\begin{equation}
N_1\leq N_3\,.
\end{equation}
Next, we can add one by one the $N_2$ lines corresponding to $\mathbf{v}$, which should go from south-west to north-east of the figure. This can be done in several equivalent ways, owing to the Yang-Baxter equation. We want to do so in such a way as to maximise the triple scattering discussed above, which here results in a triangular vertex (see also the figure). It is convenient to distinguish three cases, depending on the value of $N_2$.

\paragraph{Case~I.}
Looking at the leftmost panel in figure~\ref{fig:triangle_diagram}, we imagine adding a single diagonal line. Clearly we obtain at most as many triangles as we have horizontal lines, \textit{i.e.}\ $N_1$ triangles --- recall that $N_1\leq N_3$. We can go on adding diagonal lines as long as we saturate the purple area in the region; indeed we can do so $N_3-N_2+1$ times at most. Therefore, we have that
\begin{equation}
T(N_1,N_2,N_3)= N_2\,N_1,\qquad N_2\leq N_3-N_1+1\,.
\end{equation}

\paragraph{Case~II.}
Let us keep adding diagonal lines, looking this time at the middle panel of figure~\ref{fig:triangle_diagram}. To maximise the number of triangles, we add diagonal lines just above or just below those we had drawn above. We have two choices with $(N_1-1)$ triangles (above and below), two choices with $(N_1-2)$ triangles, and so on. Adding one line at the time, we get the sequence
\begin{equation}
\label{eq:sequence}
N_1-1,N_1-1,N_1-2,N_1-2, N_1-3,N_1-3,\dots\,,
\end{equation}
which reaches zero in $2N_1$ steps. Therefore, as long as $N_2\leq N_3-N_1+1+2N_1= N_3+N_1+1$, the number of new triangles (in the yellow region of the figure) is given by the sum
\begin{equation}
\label{eq:sumseq}
\sum_{j=1}^{K}\Big[N_1-\frac{(2j+1-(-1)^j)}{4}\Big],
\end{equation}
which can run up at most $K=N_2-(N_3-N_1+1)$ steps. The slightly quirky summand just reproduces the sequence~\eqref{eq:sequence}. Bearing in mind that we have $N_1(N_3-N_1+1)$ triangles in the purple region, with a little algebra we find that the total number of triangles is at most
\begin{equation}
\begin{aligned}
T(N_1,N_2,N_3)=\frac{N_1N_2+N_1N_3+N_1N_2}{2}-\sum_{i=1}^3\frac{N_i^2}{4}+\frac{1+(-1)^{\sum_i N_i}}{8},\\
\text{for}\quad N_3-N_1+1<N_2\leq N_3+N_1+1\,.
\end{aligned}
\end{equation}
We shall later comment on the fractional coefficients; for the moment, suffice is to say that by construction this number is an integer despite of its complicated appearance.

\paragraph{Case~III.}
Finally, we can keep adding diagonal lines like in the rightmost panel of figure~\ref{fig:triangle_diagram}. This does not generate any new triangles. Therefore, we can just use the previous estimate, taking the limit of the sum~\eqref{eq:sumseq} to be the maximal allowed value $M=2N_1$. Then we find that
\begin{equation}
T(N_1,N_2,N_3)=N_3\,N_1,\quad N_3+N_1+1<N_2\,,
\end{equation}
as we could have expected from the diagrammatical representation: we simply have one triangle for each vertex between horizontal and vertical lines.

\subsection{Improved estimate of 't~Hooft-coupling order}
\label{sec:mirror}
Armed with these estimates, it is not difficult to see case by case at which order $O(g^p)$ a digram with $(N_1,N_2,N_3)$ mirror particles contributes.

\paragraph{Case~I.} Here we have
\begin{equation}
p=\sum_{i=1}^3 N_i \ell_i +(N_3-N_1-N_2)^2,\qquad N_2\leq N_3-N_1+1\,,
\end{equation}
which is positive semi-definite regardless of the values of $\ell_i$.
\paragraph{Case~II.} In the second case we are left with
\begin{equation}
p=\sum_{i=1}^3 N_i \ell_i +\frac{1+(-1)^{\sum_i N_i}}{2},\qquad N_3-N_1+1<N_2\leq N_3+N_1+1\,.
\end{equation}
Note that the fact that $p$ is not necessarily even is due to the fact that we are considering \textit{a single} hexagon, and have split the measure and bridge length as in eq.~\eqref{eq:hexagonschematic}. At weak coupling, we expect the expansion to be in powers of~$g^2$ when all hexagons are glued together.
\paragraph{Case~III.} Similar to case~I we have
\begin{equation}
p=\sum_{i=1}^3 N_i \ell_i +(N_2-N_1-N_3)^2,\qquad N_2>N_3+N_1+1\,.
\end{equation}

\paragraph{Infinite chains of mirror magnons.}
Notice that while the present estimate makes~$p$ bounded from below, unlike the orginal na\"ive estimate~\eqref{eq:pnaive}, it is still possible to have infinitely many processes contributing at a given order. This can be achieved by setting \textit{e.g.} $N_3$ to a fixed values and varying $N_1$ and $N_2$, \textit{provided that both bridge lengths} $\ell_1, \ell_2$ \textit{vanish}. More generally, it is only possible for infinitely many magnons to contribute at a given order of the small-coupling expansion of a single hexagon when two bridge lengths vanish. This situation may well appear when computing correlators, and is all the more frequent when considering non-planar topologies, see \textit{e.g.}\ ref.~\cite{Eden:2017ozn}. We could not find a way to rule out such processes on general grounds, though it is possible to make some progress when considering particular correlators. In section~\ref{sec:comments} we will look in detail at how these processes appear in the four-point functions of BPS operators. We will see that it is still possible to use the bound we have obtained here, along with some observations from field theory, to reduce the computation to a \textit{finite} set of mirror exchanges.

\subsection{Physical magnons}
\label{sec:physical}

From eq.~\eqref{eq:phys-virtual} we see that, as long as physical magnons are given by fundamental \textit{fermionic} excitations --- for instance, for non-BPS operators in the $sl(2)$ sector --- our estimates from the previous section apply immediately since no negative powers of~$g$ emerge from physical-mirror scattering. Things are a little more involved if we allow for bosonic excitations. Consider for instance a hexagon containing a set of physical particles $\{\mathbf{x}\}$ in the $so(6)$ sector, so that the hexagon is schematically
\begin{align}
\label{eq:hexagonschematicphys}
\sqrt{\mu_{\mathbf{u}}\mu_{\mathbf{v}}\mu_{\mathbf{w}}}\,e^{\frac{i}{2} \tilde{E}_{\mathbf{u}}\ell_{\mathbf{u}}+\dots}\,e^{i \psi \mathbf{J}+\dots}\,
\langle \,\mathfrak{h}\,|\emptyset,\mathbf{u}^{3\gamma},\emptyset,\mathbf{v}^{\gamma},\mathbf{x},\mathbf{w}^{-\gamma}\rangle = O(g^{p})\,.
\end{align}
The contribution of the $M=|\mathbf{x}|$ physical $so(6)$ magnons na\"ively would be
\begin{equation}
p\to p - M\,N_2\qquad\text{na\"ively},
\end{equation}
 which once again would yield an  estimate which is unbounded from below as $N_2$~grows. Again we can improve our estimate, this time using the triangle identity~\eqref{eq:trianglephys}.

\begin{figure}[t]
\centering
\includegraphics[width=\textwidth]{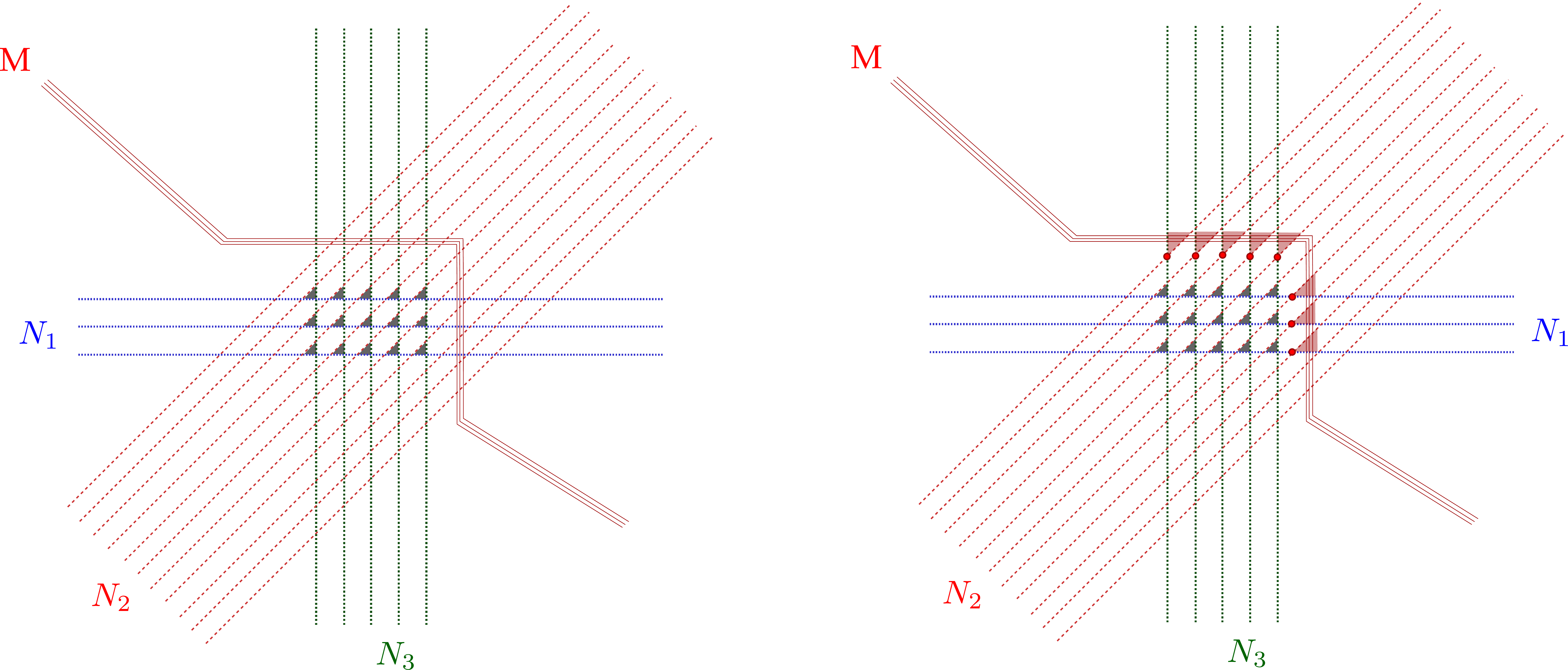}
\caption{
We draw again the scattering processes of figure~\ref{fig:triangle_diagram}, focussing on case~III and adding~$M$ physical magnons. To guide the eye, we have dotted the lines corresponding to mirror magnons. On the left, we have $M\,N_2$ scattering events between physical magnons and magnons on the opposing edge.
On the right panel, we highlight the ``cones'' emanating from the scattering of a mirror particle on the opposite edge with one from the adjacent edge; this is the vertex of the cone, and it is marked by a dot. There are clearly $N_1+N_3$ such cones.
}
\label{fig:physical}
\end{figure}

\paragraph*{Improved estimate.}
Consider the picture in figure~\ref{fig:physical}. Here we have further decorated a graph such as the one of figure~\ref{fig:triangle_diagram} by adding a new set of lines, denoted by~$M$, and corresponding to $M$~physical magnons.
We are focussing here on what we called case~III in figure~\ref{fig:triangle_diagram}, which happens when
\begin{equation}
\label{eq:caseiii}
N_2>N_1+N_3+1\,.
\end{equation}
 The reason is that the only runaway behaviour can arise as $N_2\to\infty$, which constrains us to this scenario.
We have already maximised the number of $u$-$v$-$w$ triangles following the logic explained above. Now we want to maximise the number of multiple scattering events as in eq.~\eqref{eq:trianglephys}. These can be thought of as ``cones'' emanating either from a $u$-$v$ triangle or $v$-$w$ vertex, and involving an arbitrary number of physical particles~$x$. From the figure we see that there are $N_1$ $u$-$v$ vertices and $N_3$  $v$-$w$ vertices, bearing~\eqref{eq:caseiii} in mind, for a total of $N_1+N_3$ cones. Furthermore, each cone contains~$M$ scattering events between physical magnons and mirror magnons on the opposite edge. Hence, our na\"ive estimate is improved by a factor of $M(N_1+N_3)$. All in all,
\begin{equation}
p\to p - M\,(N_2-N_1-N_3)\,,
\end{equation}
so that in presence of~$M$ physical magnons
\begin{equation}
p=\sum_{i=1}^3N_i\ell_i+(N_2-N_1-N_3)(N_2-N_1-N_3-M)\,.
\end{equation}
Hence $p$ is still bounded from below, even if its minimum value (as $N_2$ varies) can become negative. In particular, we find that
\begin{equation}
p_{\text{min}}=N_1(\ell_1+\ell_2)+N_3(\ell_3+\ell_2) -\frac{M^2-2M}{4}\,,
\end{equation}
which is bounded from below for fixed~$M$ (\textit{i.e.}, for any given physical state).%
\footnote{%
It is also worth noticing that the length~$L$ of a physical operator containing $M$ $so(6)$ excitations is $L\geq M$. This means that, for the adjacent mirror edges, $\ell_1+\ell_3\geq M$, which further improves our estimate.
}
 Similar bounds can be derived for more general configurations of physical magnons, see appendix~\ref{app:physical}.

\paragraph{Application to
correlators of non-BPS operators.}
Our estimate for~$p$ remains bounded from below and is only slightly worsened by the inclusion of physical magnons if those sit in the $so(6)$~sector. Moreover, we might expect the estimate to further improve when restricting to a particular topology for the correlator --- as it is the case for three-point functions~\cite{Basso:2015eqa}. This could arise both from glueing different hexagons, from summing over the partitions of physical magnons, and from exploiting the fact that physical states satisfy the Bethe ansatz equations.
It is still a bit surprising that $so(6)$ physical excitations behave so differently from other magnons in the physical-mirror scattering. This is essentially due to ``string-frame'' factors~\cite{Arutyunov:2006yd,Arutyunov:2009ga} in the $su(2|2)$ scattering matrix, which in the mirror kinematics affects the $g$-scaling. Historically, the hexagon formalism has been formulated in a hybrid of the spin-chain and string frames~\cite{Basso:2015zoa}, where for instance crossing transformations are performed in the string frame, but edge-widths are written in terms of spin-chain lengths; it is also apparently necessary to explicitly keep track of frame factors in certain computations~\cite{Fleury:2016ykk,Fleury:2017eph}.
As those computations have so far focussed on non-protected operators in the $sl(2)$ sector only, it would be very interesting to derive correlation functions of operators from the $su(2)$ subsector via the hexagon approach, and understand how these match the field-theory results.

\begin{figure}[t]
\centering
\includegraphics[width=0.8\textwidth]{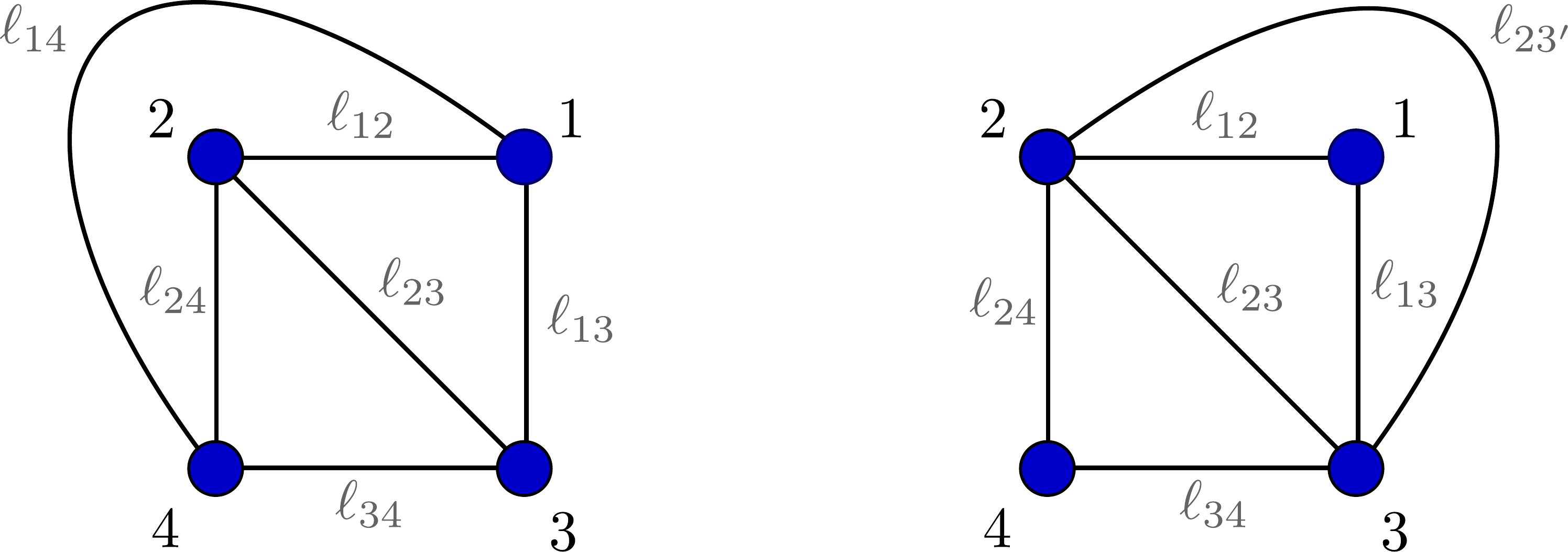}
\caption{
Hexagon tessellations for planar four-point functions. The four points lie on a sphere which we represent on the plane. Solid lines represent strands of propagators, see eq.~\eqref{eq:propagators}, which dictate in which way to tessellate the diagram. Up to relabeling the points $1234$ these are the only planar tree-level diagrams~\cite{Eden:2016xvg,Fleury:2016ykk}. Planarity restricts the $\ell_{ij}$ that can be non-zero to those in the figure. Still, it is possible that a subset of the solid lines in the figure have $\ell_{ij}=0$, as we shall see~below.
}
\label{fig:fourpt}
\end{figure}

\section{Application to planar BPS four-point functions}
\label{sec:comments}

In this section we apply our results above to the computation of a particular correlation function; we will especially focus on the issue of infinite chains of mirror magnons. Let us consider a planar four-point function of half-BPS operators
\begin{equation}
\langle \, \cO_{k_1}(x_1) \, \cO_{k_2}(x_2) \,
\cO_{k_3}(x_3) \,\cO_{k_4}(x_4) \, \rangle \, , \qquad \cO_k \, = \,
\mathrm{Tr}(\hat Z^k)\,,
\end{equation}
where the six real scalars $\phi^I$ of $\cN = 4$ SYM are rewritten in the linear combination $\hat Z = \sum_{I=1}^6\phi^I Y^I$, with $Y$ a complex null vector. Two fields $\hat{Z}_i(x_i)$ and $\hat{Z}_j(x_j)$ have the free propagator
\begin{equation}
\Pi_{ij}= \frac{Y_i\cdot Y_j}{(x_i-x_j)^2}\,.
\end{equation}

As proposed in refs.~\cite{Eden:2016xvg,Fleury:2016ykk}, to compute such a four-point function by hexagon tessellations we start by listing all propagator combinations with the right conformal weights $k_1,k_2,k_3,k_4$. These are simply
\begin{equation}
\label{eq:propagators}
\Bigl\{\Pi_{12}^{\ell_{12}} \, \Pi_{13}^{\ell_{13}} \, \Pi_{14}^{\ell_{14}} \,
\Pi_{23}^{\ell_{23}} \, \Pi_{24}^{\ell_{24}} \, \Pi_{34}^{\ell_{34}} \, : \quad k_i
= \sum_j \ell_{ij} \, \Bigr\} \,.
\end{equation}
Notice that the number of propagator $\ell_{ij}$ in the strand going from point~$i$ to point~$j$ is exactly the \textit{bridge length} which we discussed in section~\ref{sec:positivity1}, and which is a crucial part of our final estimate for $p$, see eq.~\eqref{eq:pcomplete} and below.
We are interested in planar graphs, \textit{i.e.}\ graphs that can be drawn on a sphere without intersecting propagators; not all assignments of $\ell_{ij}$ give rise to such graphs. It is easy to see this diagrammatically. In figure~\ref{fig:fourpt} we show the choices of $\ell_{ij}$ that give rise to planar graphs and the relative hexagon tessellation. For any allowed choice of propagators resulting in such a planar topology from the set~\eqref{eq:propagators}, we have to decorate the figure with all mirror magnons allowed by our estimates of section~\ref{sec:positivity1}, up to the order in~$g^2$ in which we are interested.

\begin{figure}[t]
\centering
\includegraphics[width=0.8\textwidth]{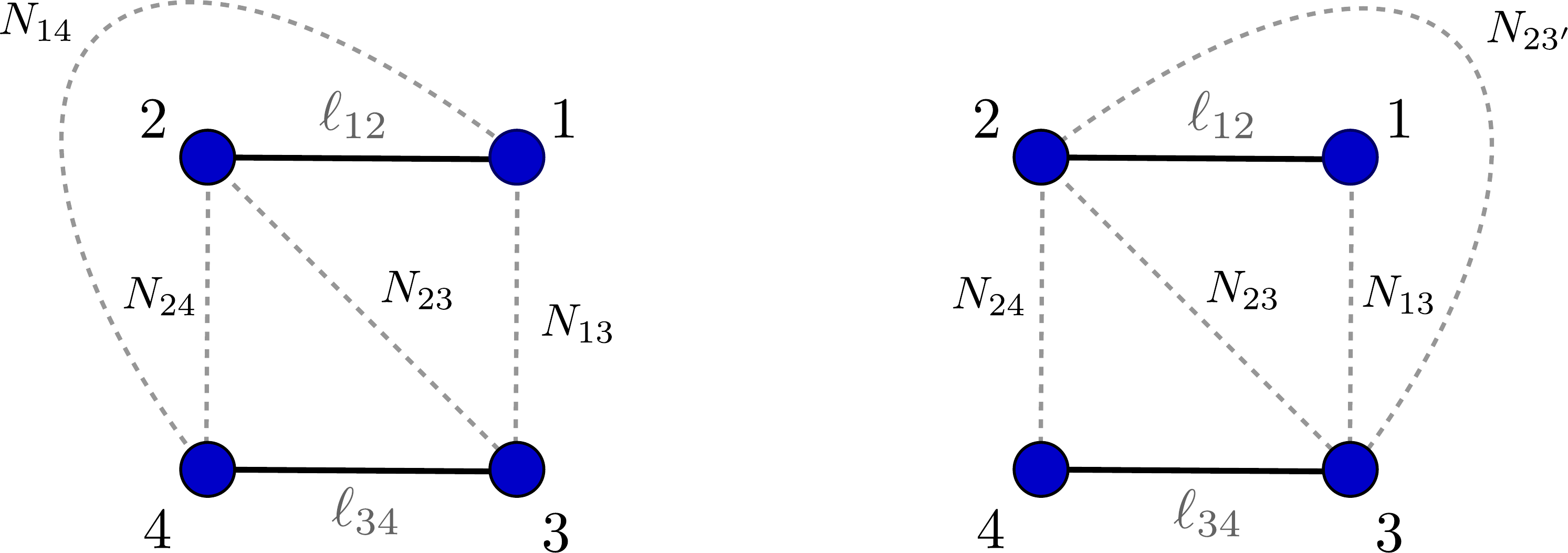}
\caption{Four-point functions with four vanishing bridge lengths.
The resulting hexagon tessellations allow for infinite chains of mirror magnons (along the zero-length bridges) already at order~$g^0$.}
\label{fig:fourptzerolength}
\end{figure}

Notice that already at the lowest orders in $g\ll1$ this can involve an \textit{infinite chain of mirror particles}. More specifically, this can happen when sufficiently many mirror edges have vanishing bridge length. Consider figure~\ref{fig:fourptzerolength}, and notice that \textit{e.g.}\ in the left diagram
\begin{equation}
\label{eq:lcondition}
\ell_{24}=\ell_{23}=\ell_{13}=\ell_{14}=0\,.
\end{equation}
Let us now estimate the order~$g^p$ at which this graph contriutes,  restricting to the case where there are no mirror magnons on the non-zero length bridges ($N_{12}=N_{34}=0$), which turns out to be the key example. Then we have%
\footnote{Notice that here there is no possibility to have ``triple scattering events'' (see section~\ref{sec:positivity1}) and our estimate for the 't-Hooft coupling order~$g^p$ coincides with the na\"ive one~\eqref{eq:pnaive}.}
\begin{equation}
\label{eq:typeIchain}
\begin{aligned}
p&=\ 2\,(N_{24}^2+N_{23}^2+N_{13}^2+N_{14}^2-N_{24}N_{23}-N_{23}N_{13} -N_{13}N_{14}-N_{12}N_{24})\\
&=\ (N_{24}-N_{23})^2+(N_{23}-N_{13})^2+(N_{13}-N_{14})^2+(N_{14}-N_{24})^2\,.
\end{aligned}
\end{equation}
In the first line, the quadratic contributions come from the interaction of magnons on the same mirror edge (including the mirror measure), while the bilinear terms come from scattering magnons on different edges within the same hexagon. The expression on the second line makes it transparent that we can have infinitely many mirror magnons contributing at given orders of perturbation theory~$g^p$ by tuning the values of $N_{ij}$.
For instance, this happens when $N_{24}=N_{23}=N_{13}=N_{14}=N$ and we let $N=1,2,\dots$. This process should contribute \textit{already at tree-level!} Going to slightly higher orders in~$g^2$, more trouble appears. Take for instance $N=N_{24}=N_{23}=N_{13}=N_{14}-1$: all these processes would appear at order~$g^2$; similarly, we find more such ``chains'' at higher orders. Moreover, depending on the loop order which interests us we may also need to consider more complicated processes. For instance, by putting a single mirror magnon on edge~12 ($N_{12}=1$) we obtain
\begin{equation}
\label{eq:typeIIchain}
p=2\ell_{12}+(N_{24}-N_{23})^2+(N_{23}-N_{13}+1)^2+(N_{13}-N_{14})^2+(N_{14}-N_{24}+1)^2,
\end{equation}
where it was crucial to use the improved estimate of section~\ref{sec:positivity1} for~$p$ to be positive. We see that such processes give infinite chains of magnons starting from order~$g^{2\ell_{12}}$. Clearly the same can be done on edge~34.

What is the physical interpretation of these processes? Since they only stem from disconnected tree-level graphs, they seem quite pathological; however, it is not obvious that we can discard them.
The tree-level graph should indeed be trivial, so that the~$g^0$ contribution from this infinite chain of magnons vanish. However, at higher orders, we might imagine that introducing mirror particles corresponds to inserting virtual gluon~lines in field theory~\cite{Eden:2017ozn}, which might give a connected four-point graph.
At one loop, it has been shown that the hexagon prediction \textit{without any infinite chain} matches the field theory result~\cite{Fleury:2016ykk}, so it seems that these vanish at that order too. What about higher loops?

As such infinite chains can hardly be evaluated directly, to address this problem it is useful to notice that there is some redundancy in the hexagon formulation, which results in linear identities among apparently unrelated mirror-magnon processes. In particular, we note the following conditions
\begin{enumerate}
\item When several $\ell_{ij}$s vanish it is possible to embed the same tree-level graph into different tessellations. It is natural to assume that all such embeddings are equivalent. This condition, which we call \emph{embedding invariance}, has been checked on several examples so far~\cite{Eden:2016xvg,Fleury:2016ykk,Eden:2017ozn}.
\item The hexagon formula~\eqref{eq:hexagonschematicmirror} depends only on the lengths of the bridges which contain at least one mirror magnon~\cite{Fleury:2016ykk}; hence diagrams with no magnons on edge $ij$ give identical integrals for any~$\ell_{ij}$. We call this property \emph{forgetfulness} of the hexagons.
\item In $\mathcal{N}=4$ SYM there exist \emph{magic identities} for four point functions~\cite{Drummond:2006rz}, due to superconformal symmetry.
\item It is also known that certain four-point functions obey \emph{non-renormalisation} theorems~\cite{DHoker:1999jke,Bianchi:1999ie,Eden:1999kw}; this happens for of ``extremal'' and ``sub-extremal'' correlators, in which the lengths of the four operators obey $L_1 = L_2 + L_3 + L_4$ and $L_1 = L_2 + L_3 + L_4 - 2$, respectively.
\item Furthermore, we can match correlators on \emph{explicit results} \cite{Chicherin:2015edu} for the $SU(N)$ gauge group; this gives constraints for the mirror-magnon processes that appear in more general situations, too.
\end{enumerate}
Using these criteria, we see\footnote{%
We have employed all possible correlators of half-BPS operators with the length of each of the four operators ranging from~2 to~7.}
 that infinite chains of mirror magnons of the type of eq.~\eqref{eq:typeIIchain} \textit{vanish} also at order~$g^4$ (we might have expected such processes to appear when $\ell_{12}=2$ or $\ell_{34}=2$). As for the chain of eq.~\eqref{eq:typeIchain}, it can be recast as a sum of finitely many mirror processes (involving at most three mirror magnons), so that it can be evaluated directly.
It is of course somewhat disappointing that these considerations, much like the ones at order~$g^2$ and~$g^0$ do not stem only from the internal consistency of the hexagon formalism, but require input from field theory.

\section{Conclusions and outlook}
\label{sec:conclusions}
In this paper we have improved on the na\"ive bound for the 't~Hooft-coupling scaling of a hexagon form factor with a set of $N_1,N_2,N_3$ mirror magnons. That would have been order~$g^p$ with
\begin{equation}
p_{\text{na\"ive}}=\sum_{i=1}^3N_i\ell_i+(N_1-N_2)^2+N_3^2-2N_3(N_1+N_2)\,,
\end{equation}
which is unbounded from below as $N_1, N_2$ and $N_3$ grow --- signalling an apparent breakdown of the weak-coupling perturbation theory for the hexagon form-factor program and flying in the face of established perturbative results such as those of refs.~\cite{Eden:2016xvg,Fleury:2016ykk,Eden:2017ozn,Fleury:2017eph}. We have shown that such na\"ive bound can be improved and is non-decreasing as $N_1, N_2$ and $N_3$ grow. Similar considerations apply in the presence of physical magnons.

However, even our improved formula allows for infinite ``chains'' of mirror particles at finite order in~$g^p$. They appear only when two out of three bridge lengths~$\ell_i$ vanish on a given hexagon. In the case of four-point functions, which we investigated at some length in section~\ref{sec:comments}, this is what happens for tessellations built from \textit{disconnected} tree-level graphs. While we could not find an argument to argue these chains away within the hexagon formalism, we have observed that in practice (taking into account field theory considerations) we do not need to compute such an infinite sum of terms; we can instead trade them for a finite sum of mirror-magnon contributions, at least up two loops in field theory --- order~$g^4$.
It is not hard to believe that such infinite chains of magnons might simplify drastically: viewing the resulting sum-integrals as Mellin representations~\cite{Fleury:2016ykk} one might \textit{e.g.}\ expect that a chain of gluings over zero-width edges can be simplified as in Barnes' lemma.\footnote{B.E.\ thanks Benjamin Basso for a discussion on related matters.}
Still it would be very interesting to understand from first principles whether such infinite chains of magnons should be disregarded and why.

Having a systematic understanding of the dynamics of mirror magnons is an important first step towards the summation of finite-size corrections to obtain a truly non-perturbative framework, perhaps along the lines of the mirror thermodynamic Bethe ansatz formalism for two-point functions. It is not obvious what is the best context to tackle this important and challenging problem. While the hexagon formalism is established only for AdS$_5/$CFT$_4$, where the integrability machinery is best developed, it is interesting to note that in integrable AdS$_3/$CFT$_2$~\cite{Sfondrini:2014via} there exist models where wrapping effects drastically simplify~\cite{Baggio:2018gct,Dei:2018mfl} --- in fact, they \textit{vanish} for two-point functions --- and where closed formulae exist for some correlation functions owing to worldsheet (Wess-Zumino-Witten) techniques. This might also be an excellent arena to test the hexagon~program.

Let us also remark that the same reasoning that allowed us to circumvent infinite sums of mirror magnons (as explained in section~\ref{sec:comments}) also leads us to a number of interesting observations which would greatly simplify the computation of the four-BPS correlator at order~$g^4$.
For instance, a rather problematic  class of diagrams is the one which we can obtain from the disconnected ones of figure~\ref{fig:fourptzerolength} by setting \textit{e.g.}\ $\ell_{13}=1$; we call these ``sausage'' graphs. There, infinite chains of mirror magnons do not appear but we find a large number of possible processes: at order~$g^2$ we should consider up to \textit{four} mirror magnons, while at order~$g^4$ we should consider up to \textit{eight} mirror magnons. Such a proliferation is worrying, as it makes the hexagon formalism very cumbersome. On the other hand, using the criteria 1.--5.\ of the section above, we can see that there are numerous cancellations, which happen graph-by-graph for processes involving different numbers of mirror magnons. In particular, processes involving four magnons are suppressed and only appear at order~$g^4$, and processes involving five or more magnons only appear at order~$g^6$.
It is interesting to observe that this cancellation is also compatible with a further field-theory inspired constraint, that is the \textit{maximum transcendentality} of a correlator increases with the order of~$g^2$. Technically, this order appears in the hexagon formalism from the highest-pole order in the integrand of mirror processes. This pole comes%
\footnote{%
This can be seen most readily by rewriting the integrand by taking partial fractions, like in refs.~\cite{Eden:2015ija,Basso:2015eqa}, and observing that partial fractioning preserves the order of the leading pole.
}
from the mirror measure and bridge-length contribution. On these grounds, we would expect that mirror integrals related to edges with length~$\ell$ only contribute at order~$g^{2\ell+2}$ or higher, as it is indeed the case for sausage graphs. In fact, this principle may be taken as an additional constraint in future hexagon-form-factor computations.

\subsection*{Acknowledgements}

BE is supported by DFG ``eigene Stelle'' Ed 78/4-3. MdL was supported by SFI and the Royal Society for funding under grant UF160578. D.~le Plat acknowledges support by the Stiftung der Deutschen Wirtschaft.
AS's work is partially supported by the NCCR SwissMAP, funded by the Swiss National Science Foundation; he also acknowledges support by the ETH ``Career Seed Grant'' no.~0-20313-17.

\appendix

\section{Algorithm for triple mirror scattering}
\label{sec:positivity2}
\begin{figure}[t]
\begin{center}
\includegraphics[scale=0.3]{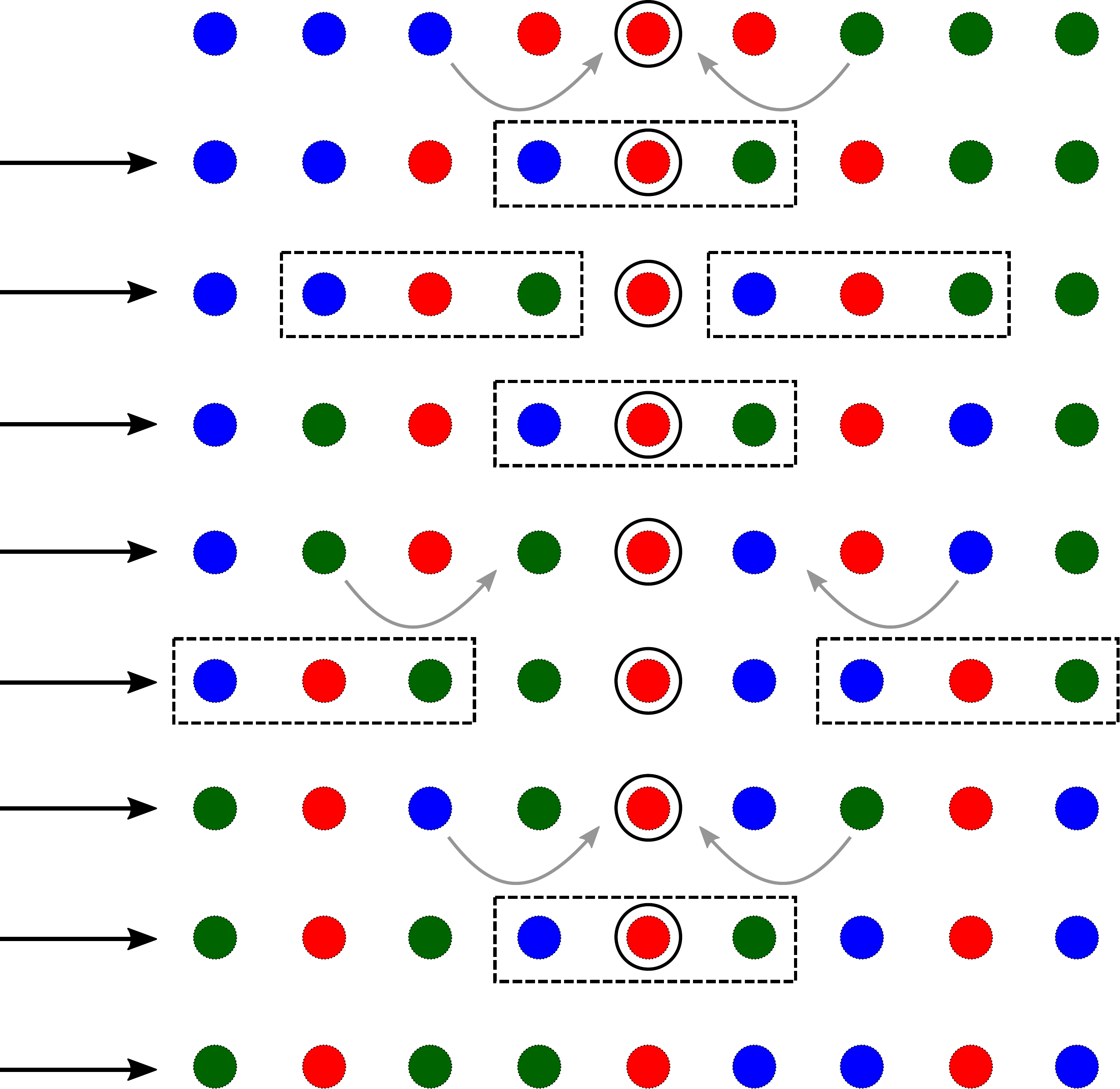}
\caption{Diagrammatic representation for the scattering processes. The blue, red and green dots stand for particles \one,\two and \three respectively. The circled red dot denotes the seed particle \textit{\two} and the dashed box represent the triple scattering process [\one\two\three]. We see 7 triple scattering processes in the diagram.}
\label{fig:example}
\end{center}
\end{figure}

In this section we give an alternative derivation of the number of triple scattering events ${T}(N_1,N_2,N_3)$ discussed in section~\ref{sec:positivity1}. This derivation is based on an algorithmic derivation of the scattering process that maximize the number of triangles. In order to introduce the notation and explain the idea, we first consider a simple example with $(N_1,N_2,N_3)=(3,3,3)$. In what follows, we will denote the configuration $(N_1,N_2,N_3)$ by
\begin{align}
\underbrace{\textcolor{blue}{1\cdots1}}_{N_1}\underbrace{\textcolor{red}{2\cdots2}}_{N_2}\underbrace{\textcolor{darkgreen}{3\cdots3}}_{N_3}
\equiv(\textcolor{blue}{1})^{N_1}(\textcolor{red}{2})^{N_2}(\textcolor{darkgreen}{3})^{N_3}
\end{align}
We expect to have ${T}(3,3,3)=7$ from the previous section. Here we demonstrate how this can be realized. We start with the configuration
\begin{align}
\one\one\one\two\textit{{\two}}\two\three\three\three.
\end{align}
According to the hexagon form factor prescription, we move the $\one$'s to the right and $\three$'s to the left by scattering adjacent magnons by the $S$-matrix.
The process is as follows
\begin{align}
\label{eq:example}
&\one\one\one\two\textit{{\two}}\two\three\three\three\longrightarrow \one\one\two[\one\textit{\two}\three]\two\three\three
\longrightarrow\one[\one\two\three]\textit{\two}[\one\two\three]\three\\\nonumber
\longrightarrow\,&\one\three\two[\one\textit{\two}\three]\two\one\three
\longrightarrow\one\three\two\three\textit{\two}\one\two\one\three
\longrightarrow[\one\two\three]\three\textit{\two}\one[\one\two\three]\\\nonumber
\longrightarrow\,&\three\two\one\three\textit{\two}\one\three\two\one
\longrightarrow\three\two\three[\one\textit{\two}\three]\one\two\one
\longrightarrow\three\two\three\three\textit{\two}\one\one\two\one
\end{align}
The rest of the scattering cannot involve the triple processes that we are looking for, so we stop here. In the above, the particle in the middle is denoted by italic font $\textit{\two}$. This particle is special for our algorithm and is called the \emph{seed particle}. The triple scattering process which corresponds to the triangle in the previous section is denoted by
\begin{align}
\cdots[\one\two\three]\cdots\longrightarrow\cdots\three\two\one\cdots
\end{align}
The process described in (\ref{eq:example}) can be represented diagrammatically as in figure~\ref{fig:example}.

Now we give the general proof. As before, we need to consider several cases while performing the counting. We split the scattering process into two parts. The first part is universal for all the cases. The second part depends on the various cases we consider.

\subsection{Universal move}
We start with the following configuration with the seed particle
\begin{align}
\one^{N_1}\two^{N_2}\three^{N_3}=\one^{N_1}\two^S\textit{\two}\two^{N_2-S-1}\three^{N_3},
\end{align}
where
\begin{align}
S=\left[\frac{N_1+N_2-N_3}{2}\right].
\end{align}
We move the particles \one/\three towards the seed particle from the left/right. The first step is
\begin{align}
\one^{N_1}\two^S\textit{\two}\two^{N_2-S-1}\three^{N_3}
\longrightarrow&\,\one^{N_1-1}\two^{S}[\one\textit{\two}\three]\two^{N_2-S-1}\three^{N_3-1}\\\nonumber
=&\,\one^{N_1-1}\two^{S}\three\textit{\two}\one\two^{N_2-S-1}\three^{N_3-1}\\\nonumber
=&\,\one^{N_1-1}\two^{S-1}(\two\three\!)\textit{\two}(\one\two\!)\two^{N_2-S-2}\three^{N_3-1}.
\end{align}
For the $l$-th step, we perform the move
\begin{align}
&\one^{N_1-l+1}\two^{S-l+1}(\two\three\!)^{l-1}\textit{\two}(\one\two\!)^{l-1}\two^{N_2-S-l}\three^{N_3-l+1}\\\nonumber
\longrightarrow\,&\one^{N_1-l}\two^{S-l+1}\one(\two\three\!)^{l-1}\textit{\two}(\one\two\!)^{l-1}\three\two^{N_2-S-l}\three^{N_3-l}\\\nonumber
\longrightarrow\,&\one^{N_1-l}\two^{S-l+1}(\three\two\!)^{l-1}[\one\textit{\two}\three](\two\one\!)^{l-1}\two^{N_2-S-l}\three^{N_3-l}\\\nonumber
\longrightarrow\,&\one^{N_1-l}\two^{S-l+1}(\three\two\!)^{l-1}\three\textit{\two}\one(\two\one\!)^{l-1}\two^{N_2-S-l}\three^{N_3-l}\\\nonumber
=\,&\one^{N_1-l}\two^{S-l}(\two\three\!)^{l}\textit{\two}(\one\two\!)^{l}\two^{N_2-S-l-1}\three^{N_3-l}.
\end{align}
The number of triple scattering processes involved in the $l$-th step is $2l-1$. We can perform this move until one of the exponents $N_1-l$, $S-1$, $N_2-S-l-1$ and $N_3-l$ becomes zero. Therefore we need to distinguish between different cases. First we say without loss of generality $N_2\geq N_3$ and $N_2\geq N_1$. We can always obtain this by cyclicity. We consider the case that $N_1\le N_3$. The case with $N_1>N_3$ can be dealt with in a similar way.


\subsection{The case \texorpdfstring{$S> N_1$}{S>N1} and \texorpdfstring{$N_3\geq N_1$}{N3>=N1}}
We have $N_2\ge N_1+N_3$. The universal move stops because $N_1-l$ becomes zero and we end up with the following configuration:
\begin{align}
\two^{S-N_1}(\two\three\!)^{N_1}\textit{\two}(\one\two\!)^{N_1}\two^{N_2-S-N_1-1}\three^{N_3-N_1}
\end{align}
For the rest of the argument the seed particle is not important anymore, so we treat it as the other \two particles. We perform the move
\begin{align}
&\two^{S-N_1}(\two\three\!)^{N_1}{\two}(\one\two\!)^{N_1}\two^{N_2-S-N_1-1}\three^{N_3-N_1}\\\nonumber
\longrightarrow\,&\two^{S-N_1}(\two\three\!)^{N_1}{\two}[(\one\two\!)^{N_1}\three]\two^{N_2-S-N_1-1}\three^{N_3-N_1-1}\\\nonumber
\longrightarrow\,&\two^{S-N_1}(\two\three\!)^{N_1}{\two}\three(\two\one\!)^{N_1}\two^{N_2-S-N_1-1}\three^{N_3-N_1-1}\\\nonumber
=\,&\two^{S-N_1}(\two\three\!)^{N_1+1}\two(\one\two\!)^{N_1}\two^{N_2-S-N_1-2}\three^{N_3-N_1-1}.
\end{align}
This can be done $N_3-N_1$ times. Each step involves $N_1$ triple scattering processes. Finally we end up with the configuration
\begin{align}
\two^{S-N_1}(\two\three)^{N_3}\two(\one\two)^{N_1}\two^{N_2-S-N_3-1}.
\end{align}
In this two-part procedure, we count the total number of triangle scattering processes
\begin{align}
{T}(N_1,N_2,N_3)=\sum_{l=1}^{N_1}(2l-1)+N_1(N_3-N_1)=N_1N_3,
\end{align}
which matches what we had found in section~\ref{sec:positivity1} for cases I and~III.

\subsection{The case \texorpdfstring{$S\leq N_1$}{S<=N1} and \texorpdfstring{$N_3>N_1$}{N3>N1}}
Here the universal move stops because $S-l$ becomes zero. We end up with
\begin{align}
\one^{N_1-S}(\two\three)^S\two(\one\two)^S\two^{N_2-2S-1}\three^{N_3-S}.
\end{align}
To obtain further triple scattering processes, we first perform the move
\begin{align}
&\one^{N_1-S}(\two\three)^S\two(\one\two)^S\two^{N_2-2S-1}\three^{N_3-S}\\\nonumber
=\,&\one^{N_1-S-1}[\one(\two\three)^S]\two(\one\two)^S\two^{N_2-2S-1}\three^{N_3-S}\\\nonumber
\longrightarrow\,&\one^{N_1-S-1}(\three\two)^S\one\two(\one\two)^S\two^{N_2-2S-1}\three^{N_3-S}\\\nonumber
=\,&\one^{N_1-S-1}\three(\two\three)^{S-1}\two(\one\two)^{S+1}\two^{N_2-2S-1}\three^{N_3-S}.
\end{align}
This move involves $S$ triple scattering processes. Then we perform the following recursively: at step $l$, we have
\begin{align}
&\one^{N_1-S-l}\three^l(\two\three)^{S-1}\two(\one\two)^{S+l}\two^{N_2-2S-l}\three^{N_3-S-l+1}\\\nonumber
\longrightarrow\,&\one^{N_1-S-l-1}\three^l[\one(\two\three)^{S-1}]\two[(\one\two)^{S+l}\three]\two^{N_2-2S-l}\three^{N_3-S-l}\\\nonumber
\longrightarrow\,&\one^{N_1-S-l-1}\three^l(\three\two)^{S-1}[\one\two\three](\one\two)^{S+l}\two^{N_2-2S-l}\three^{N_3-S-l}\\\nonumber
\longrightarrow\,&\one^{N_1-S-l-1}\three^l(\three\two)^{S-1}\three\two\one(\two\one)^{S+l}\two^{N_2-2S-l}\three^{N_3-S-l}\\\nonumber
=\,&\one^{N_1-S-l-1}\three^{l+1}(\two\three)^{S-1}\two(\one\two)^{S+l+1}\two^{N_2-2S-l-1}\three^{N_3-S-l}.
\end{align}
When the process ends depends on whether $N_1-S-1$ or $N_2-2S-1$ is larger. We separately consider these cases.\par
\begin{enumerate}
\item First, let $N_2-N_1\ge S$ so that the process ends because $N_1-S-l-1$ becomes zero. We end on
\begin{align}
\three^{N_1-S}(\two\three)^{S-1}\two(\one\two)^{N_1}\two^{N_2-N_1-S}\three^{N_3-N_1+1}.
\end{align}
To proceed, we perform the following recursive move: for step $l$, we have
\begin{align}
&\three^{N_1-S}(\two\three)^{S+l-2}\two(\one\two)^{N_1}\two^{N_2-N_1-S-l+1}\three^{N_3-N_1-l+2}\\\nonumber
\longrightarrow\,&\three^{N_1-S}(\two\three)^{S+l-2}\two[(\one\two)^{N_1}\three]\two^{N_2-N_1-S-l+1}\three^{N_3-N_1-l+1}\\\nonumber
\longrightarrow\,&\three^{N_1-S}(\two\three)^{S+l-2}\two\three(\two\one)^{N_1}\two^{N_2-N_1-S-l+1}\three^{N_3-N_1-l+1}\\\nonumber
=\,&\three^{N_1-S}(\two\three)^{S+l-1}\two(\one\two)^{N_1}\two^{N_2-N_1-S-l}\three^{N_3-N_1-l+1}.
\end{align}
Each step involves $N_1$ and will stop for $l=N_2-N_1-S$. So the total number of triple scattering processes is $N_1(N_2-N_1-S)$. We end up with the following configuration
\begin{align}
\three^{N_1-S}(\two\three)^{N_2-N_1-1}\two(\one\two)^{N_1}\three^{N_3-N_2+S+1}.
\end{align}
As the last step, we perform the following recursive move, for the $l$-th step
\begin{align}
&\three^{N_1-S}(\two\three)^{N_2-N_1+l-1}\two(\one\two)^{N_1-l}\one^l\three^{N_3-N_2+S-l+1}\\\nonumber
\longrightarrow\,&\three^{N_1-S}(\two\three)^{N_2-N_1+l-1}\two[(\one\two)^{N_1-l}\three]\one^l\three^{N_3-N_2+S-l}\\\nonumber
\longrightarrow\,&\three^{N_1-S}(\two\three)^{N_2-N_1+l-1}\two\three(\two\one)^{N_1-l}\one^l\three^{N_3-N_2+S-l}\\\nonumber
=\,&\three^{N_1-S}(\two\three)^{N_2-N_1+l}\two(\one\two)^{N_1-l-1}\one^{l+1}\three^{N_3-N_2+S-l}.
\end{align}
This stops at $l=N_3-N_2+S$ and end up with the following configuration
\begin{align}
\three^{N_1-S}(\two\three)^{N_3-N_1+S}\two(\one\two)^{N_1+N_2-N_3-S-1}\one^{N_3-N_2+S+1}.
\end{align}
We can not produce further triple scattering processes. Summing up the number of triple scattering processes, we obtain
\begin{align}
\label{Triplesum}
{T}(N_1,N_2,N_3)=&\,\sum_{l=1}^S(2l-1)+S+\sum_{l=1}^{N_1-S-1}(2S+l)+N_1(N_2-N_1-S)\\\nonumber
&\,+N_1+\sum_{l=1}^{N_3-N_2+S}(N_1-l)\\\nonumber
=&\,-\frac{1}{4}\left(N_1^2+N_2^2+N_3^2-2N_1N_2-2N_1N_3-2N_2N_3-c\right) , \label{Sums}
\end{align}
where $c=\text{mod}[N_1+N_2+N_3,2]$. Using the result, we find that the order $g^p$ is given by
\begin{align}
p=\sum_{i=1}^3 \ell_i N_i+\frac{1+(-1)^{\sum_i N_i}}{2}\,.
\end{align}

\item The case $N_2-N_1< S$, the process ends because $N_2-2S-1$ becomes zero. We end up with the following configuration
\begin{align}
\one^{N_1-N_2+S}\three^{N_2-2S}(\two\three)^{S-1}\two(\one\two)^{N_2-S}\three^{N_3-N_2+S+1}.
\end{align}
We now perform the following recursive move at $l$-th step. Here we get $N_2-S-l+1$ triples.
\begin{align}
&\one^{N_1-N_2+S}\three^{N_2-2S}(\two\three)^{S+l-2}\two(\one\two)^{N_2-S-l+1}\one^{l-1}\three^{N_3-N_2+S+2-l}\\
\longrightarrow&\one^{N_1-N_2+S}\three^{N_2-2S}(\two\three)^{S+l-1}\two(\one\two)^{N_2-S-l}\one^l\three^{N_3-N_2+S+1-l}.\notag
\end{align}
\end{enumerate}
This procedure stops when $N_3-N_2+S+1-l$ goes to zero. We get the following configuration:
\begin{align}
\one^{N_1-N_2+S}\three^{N_2-2S}(\two\three)^{N_3-N_2+2S}\two(\one\two)^{2N_2-2S-N_3-1}\one^{N_3-N_2+S+1}.
\end{align}
We now move all the \one from the left across all (\two\three). At $l$-th step again we get the following move:
\begin{align}
&\one^{N_1-N_2+S-l+1}\three^{N_2-2S+l-1}(\two\three)^{N_3-N_2+2S-l+1}\two(\one\two)^{2N_2-2S-N_3-l}\one^{N_3-N_2+S+1}\\
\longrightarrow&\one^{N_1-N_2+S-l}\three^{N_2-2S+l}(\two\three)^{N_3-N_2+2S-l}\two(\one\two)^{2N_2-2S-N_3-l-1}\one^{N_3-N_2+S+1}.\notag
\end{align}
By doing this step we get $N_3-N_2+2S-l+1$ new triples. This is possible until $N_1-N_2+S-l$ goes to zero. Then we end up with:
\begin{align}
\three^{N_1-S}(\two\three)^{N_3-N_1+S}\two(\one\two)^{N_2-S-N_3+N_1-1}\one^{N_3-N_2+S+1}.
\end{align}
Here we cannot produce any more triples again. Counting all triples we get the following.
\begin{align}
{T}(N_1,N_2,N_3)&=\sum\limits_{l=1}^{S}(2l-1)+S+\sum_{l=1}^{N_2-2S-1}(2S+l)+\sum_{l=1}^{N_3-N_2+S+1}(N_2-S-l+1)\notag\\
&+\sum_{l=1}^{N_1-N_2+S}(N_3-N_2+2S-l+1).
\end{align}
Simplifying this expression we obtain a result for $p$ similar to \eqref{Sums}.

\subsection{The case \texorpdfstring{$N_3=N_1$}{N3=N1} and \texorpdfstring{$S\leq N_3$}{S<=N3}}
In this case, we have again $N_2\leq N_1+N_3=2N_3$. With this constraints we get $S=\left[\frac{N_2}{2}\right]$. The universal move stops because $N_2-S-1-l$ becomes zero. We now consider to different cases.
\begin{enumerate}
\item We consider first that $N_2$ is even, then $S=\frac{N_2}{2}$.
So we end up with the following configuration:
\begin{align}
\one^{N_3-S+1}\two(\two\three)^{S-1}\two(\one\two)^{S-1}\three^{N_3-S+1}.
\end{align}
We now perform the following move and get $S-1$ triples.
\begin{align}
&\one^{N_3-S+1}\two(\two\three)^{S-1}\two(\one\two)^{S-1}\three^{N_3-S+1}\\
\longrightarrow&\one^{N_3-S}(\two\three)^{S-1}\two(\one\two)^{S}\three^{N_3-S+1}.\notag
\end{align}
Now we move recursively a \one from the left and a \three from the right simultaneously to the middle as shown in \eqref{simMove}. Here we get at the $l$-th step $2S$ new triples.
\begin{align}
\label{simMove}
&\one^{N_3-S-l+1}\three^{l-1}(\two\three)^{S-1}\two(\one\two)^{S}\one^{l-1}\three^{N_3-S+2-l}\\
=&\one^{N_3-S-l}\three^{l-1}[\one(\two\three)^{S-1}]\two[(\one\two)^{S}\three]\one^{l-1}\three^{N_3-S+1-l}\notag\\
\longrightarrow&\one^{N_3-S-l}\three^{l-1}(\three\two)^{S-1}[\one\two\three](\two\one)^{S}\one^{l-1}\three^{N_3-S+1-l}\notag\\
\longrightarrow&\one^{N_3-S-l}\three^{l}(\two\three)^{S-1}\two(\one\two)^{S}\one^{l}\three^{N_3-S+1-l}.\notag
\end{align}
This recursion ends if $N3-S-l$ becomes zero and we end up with the following:
\begin{align}
\three^{N_3-S}(\two\three)^{S-1}\two(\one\two)^{S}\one^{N3-S}\three.
\end{align}
By moving the \three on the right through the (\one\two)'s we get $S$ triples and cannot perform other triple moves then. All in all the sum of the triples is
\begin{align}
{T}(N_3,N_2,N_3)=&\sum_{l=1}^{S-1}+S-1+\sum_{l=1}^{N_3-S}(2S)+S\\
=&-\frac{N_2^2}{4}+N_2N_3.\notag
\end{align}
This is the same like in \eqref{Triplesum} with $N_1=N_3$ and $N_2$ even. So we end up with formula \eqref{Sums} for $p$.
\item Now we consider that $N_2$ is odd, then $S=\frac{N2}{2}-\frac{1}{2}$. From universal move we end up with:
\begin{align}
\one^{N_3-S}\two(\two\three)^{S}\two(\one\two)^{S}\three^{N_3-S}.
\end{align}
Now we move recursively a \one from the left and a \three from the right to the middle as shown below. At $l$-th step we get $2S+1$ triples.
\begin{align}
\label{simMove2}
&\one^{N_3-S-l+1}\three^{l-1}(\two\three)^{S}\two(\one\two)^{S}\one^{l-1}\three^{N_3-S-l+1}\\
=&\one^{N_3-S-l}\three^{l-1}[\one(\two\three)^{S}]\two[(\one\two)^{S}\three]\one^{l-1}\three^{N_3-S-l}\notag\\
\longrightarrow&\one^{N_3-S-l}\three^{l-1}(\three\two)^{S}[\one\two\three](\two\one)^{S}\one^{l-1}\three^{N_3-S-l}\notag\\
\longrightarrow&\one^{N_3-S-l}\three^{l}(\two\three)^{S}\two(\one\two)^{S}\one^{l}\three^{N_3-S-l}.\notag
\end{align}
This procedure ends if $N_3-S-l$ becomes zero. After that it is not possible to do another triple move. All in all the sum of the triples is
\begin{align}
{T}(N_1,N_2,N_3)=&\sum_{l=1}^S(2l-1)+\sum_{l=1}^{N3-S}(2S+1)\\
&=-\frac{1}{4}\left(N_2^2-4N_2N_3-1\right).
\end{align}
Which is also the same like in \eqref{Triplesum} with $N_1=N_3$ and $N_2$ odd. So we end up with formula \eqref{Sums} for $p$.
\end{enumerate}

\section{General configurations involving physical magnons}
\label{app:physical}

In the appendix we extend the discussion of section~\ref{sec:physical} by considering three families of physical magnons ${\bf x},{\bf y},{\bf z}$ as well as three families of mirror magnons ${\bf u},{\bf v},{\bf w}$, see eq.~\eqref{eq:hexagonschematic}, filling all edges of a hexagon. We label the number of mirror excitations by $N_1,N_2,N_3$ as above, while the number of physical ones is $M_1,M_2,M_3$. We take the convention that there are $N_i$ mirror magnons on the edge opposite to the one with~$M_i$ physical magnons. As discussed above, see eq.~\eqref{eq:phys-virtual}, the worst-case scenario for our estimate occurs when all physical magnons are $so(6)$ excitations. In that case, scattering mirror-physical magnons from opposite edges yields a power of~$g^{-1}$, so that we na\"ively expect the hexagon to acquire an order~$g^p$, where the contribution of physical magnons shifts the results of section~\ref{sec:mirror} as
\begin{equation}
p\to p -\sum_{i=1}^3 M_i\,N_i\,,\qquad\text{na\"ively}.
\end{equation}
However, as discussed in the main text, certain multiple scattering events scale better than predicted by~\eqref{eq:phys-virtual}. Looking at eq.~\eqref{eq:trianglephys} and recalling the discussion of section~\ref{sec:physical} we see that these configurations are cone-like sequences of mirror-physical-mirror scattering events.

\begin{figure}[t]
\centering
\includegraphics[width=\textwidth]{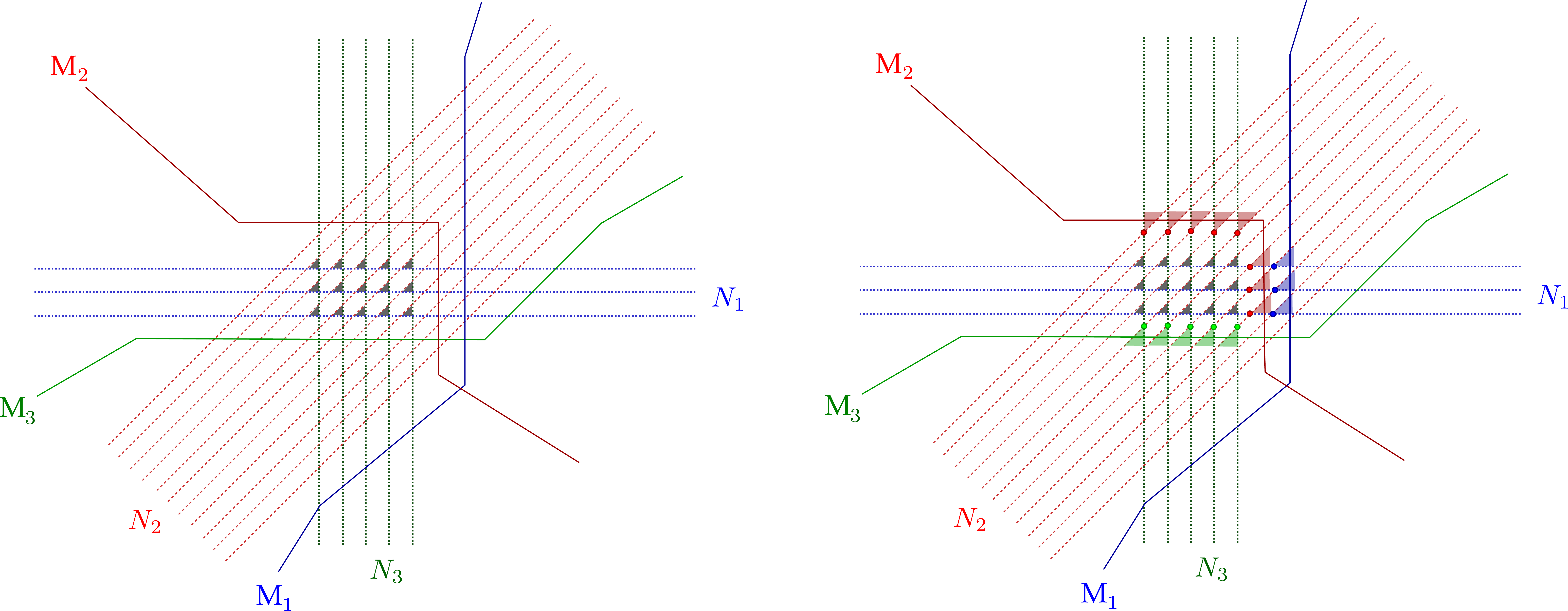}
\caption{
Scattering of physical and mirror magnons in ``case~III''. A strand of $M_i$ physical magnons is indicated by a single thick solid line, while individual mirror magnons are indicated by dashed lines. There are $M_i$ physical magnons sitting on the edge opposite to the mirror one containing $N_i$ virtual magnons.  Left panel: We count how many times physical magnons scatter with mirror ones from the opposite edge. Clearly this happens $N_i M_i$ times for $i=1,\dots 3$. Right panel: we identify the ``cones'' as in eq.~\eqref{eq:trianglephys}. In the drawing for the present case there are $N_1$ and $N_3$ such cones for the ``short'' edges, while for the ``longer'' one (here $N_2>N_1+N_3+1$) there are $(N_1+N_3)$ cones.
}
\label{fig:physicalIII}
\end{figure}

\paragraph{Estimate of cones, case III and~I.}
Let us estimate how many such ``cones'' can appear when computing the hexagon form factor. It is convenient to work diagrammatically, distinguishing the cases of the main text. We start from ``case~III'' where $N_2>N_1+N_3+1$ as discussed in section~\ref{sec:physical}, but here we consider an arbitrary numbers of physical magnons $M_1$, $M_2$ and~$M_3$, see figure~\ref{fig:physicalIII}.
We see that for the ``short'' edges $N_1, N_3$ we have $N_1$ and $N_3$ cones, while for the longer on we have $N_1+N_3$ cones. As each cone contains $M_i$ physical lines we get the improved estimate
\begin{equation}
p\to p -M_2(N_2-N_1-N_3).
\end{equation}
This happens to be the very same as the estimate for the simpler setup of section~\ref{sec:physical}, namely
\begin{equation}
p=\sum_{i=1}^3 N_i\ell_i+(N_2-N_1-N_3)(N_2-N_1-N_3-M_2)\,.
\end{equation}
 Therefore, the estimate is bounded from below as argued in the main text. Notice that the other similar cases, and in particular case~I, can be obtained by relabelling the indices $i=1,2,3$, keeping track of which $N_i$ is larger than the sum of the other two.

In figure 6 we could equally have passed the solid line representing the scattering of the $M_1$ physical excitations to the left of the central region of the panels. Then
equation (\ref{eq:trianglephys}) directly yields the aforementioned estimate because none of the cones is passed by two strands of physical magnons. The pictures in figure 6 remain correct even if the $M_1$ and $M_2$ physical magnons pass through cones with the same ``root" thanks to equation (\ref{eq216}).


\paragraph{Estimate of cones, case II.}
In this case all numbers of mirror particles $N_i$ are of the same order. More specifically, no $N_i$ is bigger than the sum of the other two. Following figure~\ref{fig:physicalII} we see that there are $(N_i-1)$ cones for any $i=1,\dots 3$, giving an improvement of $M_i(N_i-1)$. As a result, we find
\begin{equation}
p\to p -\sum_{i=1}^3 M_i,
\end{equation}
which is just an $N_i$-independent shift. Hence the final result for $p$ is in this case
\begin{equation}
p=\sum_{i=1}^3 (N_i\ell_i-M_i)+\frac{1+(-1)^{\sum_i N_i}}{2}\,.
\end{equation}

\begin{figure}[t]
\centering
\includegraphics[width=\textwidth]{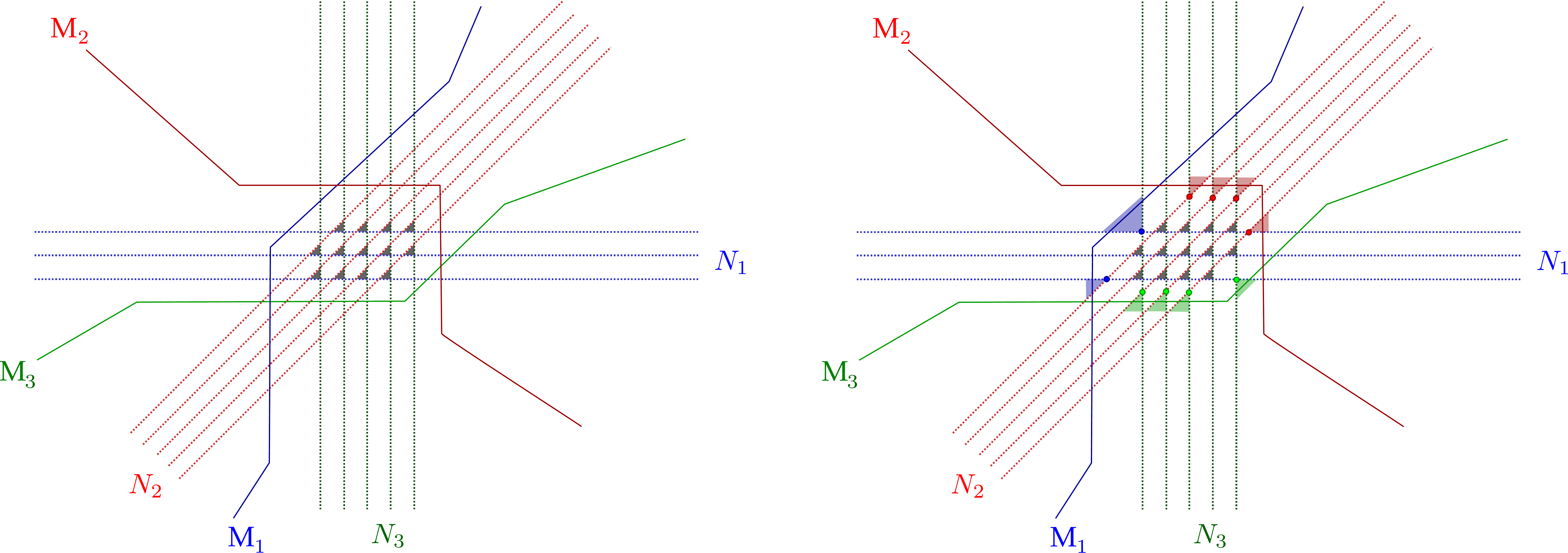}
\caption{
Scattering of physical and mirror magnons in ``case~II''. A strand of $M_i$ physical magnons is indicated by a single thick solid line, while individual mirror magnons are indicated by dashed lines (the lines are arranged differently from figure~\ref{fig:physicalIII} for clarity). There are $M_i$ physical magnons sitting on the edge opposite to the mirror one containing $N_i$ mirror magnons. Left panel: We count how many times physical magnons scatter with mirror ones from the opposite edge, which happens $N_i M_i$ times for $i=1,\dots 3$. Right panel: we identify the ``cones'' as in eq.~\eqref{eq:trianglephys}. In the drawing for this case there are $(N_i-1)$ cones for each type of lines.
}
\label{fig:physicalII}
\end{figure}


\begin{thebibliography}{10}

\bibitem{Maldacena:1997re}
J.~M. Maldacena, \emph{{The Large N limit of superconformal field theories and
  supergravity}}, \href{https://doi.org/10.1023/A:1026654312961,
  10.4310/ATMP.1998.v2.n2.a1}{\emph{Int. J. Theor. Phys.} {\bfseries 38} (1999)
  1113} [\href{https://arxiv.org/abs/hep-th/9711200}{{\ttfamily
  hep-th/9711200}}].

\bibitem{Witten:1998qj}
E.~Witten, \emph{Anti-de {S}itter space and holography}, {\emph{Adv. Theor.
  Math. Phys.} {\bfseries 2} (1998) 253}
  [\href{https://arxiv.org/abs/hep-th/9802150}{{\ttfamily hep-th/9802150}}].

\bibitem{Gubser:1998bc}
S.~S. Gubser, I.~R. Klebanov and A.~M. Polyakov, \emph{Gauge theory correlators
  from non-critical string theory},
  \href{https://doi.org/10.1016/S0370-2693(98)00377-3}{\emph{Phys. Lett.}
  {\bfseries B428} (1998) 105}
  [\href{https://arxiv.org/abs/hep-th/9802109}{{\ttfamily hep-th/9802109}}].

\bibitem{'tHooft:1973alw}
G.~'t~Hooft, \emph{A planar diagram theory for strong interactions},
  \href{https://doi.org/10.1016/0550-3213(74)90154-0}{\emph{Nucl. Phys.}
  {\bfseries B72} (1974) 461}.

\bibitem{Minahan:2002ve}
J.~A. Minahan and K.~Zarembo, \emph{The {B}ethe-ansatz for {$\superN = 4$}
  super {Y}ang-{M}ills}, {\emph{JHEP} {\bfseries 0303} (2003) 013}
  [\href{https://arxiv.org/abs/hep-th/0212208}{{\ttfamily hep-th/0212208}}].

\bibitem{Arutyunov:2009ga}
G.~Arutyunov and S.~Frolov, \emph{Foundations of the {$\AdS{5} \times S^5$}
  superstring. part {I}},
  \href{https://doi.org/10.1088/1751-8113/42/25/254003}{\emph{J. Phys. A}
  {\bfseries A42} (2009) 254003}
  [\href{https://arxiv.org/abs/0901.4937}{{\ttfamily 0901.4937}}].

\bibitem{Beisert:2010jr}
N.~Beisert et~al., \emph{Review of {AdS/CFT} integrability: An overview},
  \href{https://doi.org/10.1007/s11005-011-0529-2}{\emph{Lett. Math. Phys.}
  {\bfseries 99} (2012) 3} [\href{https://arxiv.org/abs/1012.3982}{{\ttfamily
  1012.3982}}].

\bibitem{Basso:2015zoa}
B.~Basso, S.~Komatsu and P.~Vieira, \emph{Structure constants and integrable
  bootstrap in planar {$\superN = 4$} {SYM} theory},
  \href{https://arxiv.org/abs/1505.06745}{{\ttfamily 1505.06745}}.

\bibitem{Eden:2016xvg}
B.~Eden and A.~Sfondrini, \emph{{Tessellating cushions: four-point functions in
  $\mathcal{N} $ = 4 SYM}},
  \href{https://doi.org/10.1007/JHEP10(2017)098}{\emph{JHEP} {\bfseries 10}
  (2017) 098} [\href{https://arxiv.org/abs/1611.05436}{{\ttfamily
  1611.05436}}].

\bibitem{Fleury:2016ykk}
T.~Fleury and S.~Komatsu, \emph{{Hexagonalization of Correlation Functions}},
  \href{https://doi.org/10.1007/JHEP01(2017)130}{\emph{JHEP} {\bfseries 01}
  (2017) 130} [\href{https://arxiv.org/abs/1611.05577}{{\ttfamily
  1611.05577}}].

\bibitem{Eden:2017ozn}
B.~Eden, Y.~Jiang, D.~le~Plat and A.~Sfondrini, \emph{{Colour-dressed hexagon
  tessellations for correlation functions and non-planar corrections}},
  \href{https://doi.org/10.1007/JHEP02(2018)170}{\emph{JHEP} {\bfseries 02}
  (2018) 170} [\href{https://arxiv.org/abs/1710.10212}{{\ttfamily
  1710.10212}}].

\bibitem{Bargheer:2017nne}
T.~Bargheer, J.~Caetano, T.~Fleury, S.~Komatsu and P.~Vieira, \emph{{Handling
  Handles I: Nonplanar Integrability}},
  \href{https://arxiv.org/abs/1711.05326}{{\ttfamily 1711.05326}}.

\bibitem{Carlson:2011hy}
W.~Carlson, R.~de~Mello~Koch and H.~Lin, \emph{{Nonplanar Integrability}},
  \href{https://doi.org/10.1007/JHEP03(2011)105}{\emph{JHEP} {\bfseries 03}
  (2011) 105} [\href{https://arxiv.org/abs/1101.5404}{{\ttfamily 1101.5404}}].

\bibitem{deMelloKoch:2018tlb}
R.~de~Mello~Koch, M.~Kim and H.~J.~R. Zyl, \emph{{Integrable Subsectors from
  Holography}}, \href{https://doi.org/10.1007/JHEP05(2018)198}{\emph{JHEP}
  {\bfseries 05} (2018) 198}
  [\href{https://arxiv.org/abs/1802.01367}{{\ttfamily 1802.01367}}].

\bibitem{Kristjansen:2010kg}
C.~Kristjansen, \emph{Review of {AdS/CFT} integrability, {C}hapter {IV.1}:
  Aspects of non-planarity},
  \href{https://doi.org/10.1007/s11005-011-0514-9}{\emph{Lett. Math. Phys.}
  {\bfseries 99} (2010) 349} [\href{https://arxiv.org/abs/1012.3997}{{\ttfamily
  1012.3997}}].

\bibitem{Beisert:2005fw}
N.~Beisert and M.~Staudacher, \emph{Long-range {$\grpPSU(2,2|4)$} {B}ethe
  ansaetze for gauge theory and strings},
  \href{https://doi.org/10.1016/j.nuclphysb.2005.06.038}{\emph{Nucl. Phys.}
  {\bfseries B727} (2005) 1}
  [\href{https://arxiv.org/abs/hep-th/0504190}{{\ttfamily hep-th/0504190}}].

\bibitem{Ambjorn:2005wa}
J.~Ambj{\o}rn, R.~A. Janik and C.~Kristjansen, \emph{Wrapping interactions and
  a new source of corrections to the spin-chain/string duality},
  \href{https://doi.org/10.1016/j.nuclphysb.2005.12.007}{\emph{Nucl. Phys.}
  {\bfseries B736} (2006) 288}
  [\href{https://arxiv.org/abs/hep-th/0510171}{{\ttfamily hep-th/0510171}}].

\bibitem{Luscher:1985dn}
M.~L{\"u}scher, \emph{Volume dependence of the energy spectrum in massive
  quantum field theories. 1. {S}table particle states},
  \href{https://doi.org/10.1007/BF01211589}{\emph{Commun. Math. Phys.}
  {\bfseries 104} (1986) 177}.

\bibitem{Luscher:1986pf}
M.~L{\"u}scher, \emph{Volume dependence of the energy spectrum in massive
  quantum field theories. 2. {S}cattering states},
  \href{https://doi.org/10.1007/BF01211097}{\emph{Commun. Math. Phys.}
  {\bfseries 105} (1986) 153}.

\bibitem{Eden:2015ija}
B.~Eden and A.~Sfondrini, \emph{{Three-point functions in ${\cal N}=4$ SYM: the
  hexagon proposal at three loops}},
  \href{https://doi.org/10.1007/JHEP02(2016)165}{\emph{JHEP} {\bfseries 02}
  (2016) 165} [\href{https://arxiv.org/abs/1510.01242}{{\ttfamily
  1510.01242}}].

\bibitem{Basso:2015eqa}
B.~Basso, V.~Goncalves, S.~Komatsu and P.~Vieira, \emph{{Gluing Hexagons at
  Three Loops}},
  \href{https://doi.org/10.1016/j.nuclphysb.2016.04.020}{\emph{Nucl. Phys.}
  {\bfseries B907} (2016) 695}
  [\href{https://arxiv.org/abs/1510.01683}{{\ttfamily 1510.01683}}].

\bibitem{Basso:2017muf}
B.~Basso, V.~Goncalves and S.~Komatsu, \emph{{Structure constants at wrapping
  order}}, \href{https://doi.org/10.1007/JHEP05(2017)124}{\emph{JHEP}
  {\bfseries 05} (2017) 124}
  [\href{https://arxiv.org/abs/1702.02154}{{\ttfamily 1702.02154}}].

\bibitem{Fleury:2017eph}
T.~Fleury and S.~Komatsu, \emph{{Hexagonalization of Correlation Functions II:
  Two-Particle Contributions}},
  \href{https://doi.org/10.1007/JHEP02(2018)177}{\emph{JHEP} {\bfseries 02}
  (2018) 177} [\href{https://arxiv.org/abs/1711.05327}{{\ttfamily
  1711.05327}}].

\bibitem{Beisert:2005tm}
N.~Beisert, \emph{{The $SU(2|2)$ dynamic S-matrix}},
  \href{https://doi.org/10.4310/ATMP.2008.v12.n5.a1}{\emph{Adv. Theor. Math.
  Phys.} {\bfseries 12} (2008) 945}
  [\href{https://arxiv.org/abs/hep-th/0511082}{{\ttfamily hep-th/0511082}}].

\bibitem{Arutyunov:2009mi}
G.~Arutyunov, M.~de~Leeuw and A.~Torrielli, \emph{The bound state {S}-matrix
  for {$\AdS{5} \times \Sphere^5$} superstring},
  \href{https://doi.org/10.1016/j.nuclphysb.2009.03.024}{\emph{Nucl. Phys.}
  {\bfseries B819} (2009) 319}
  [\href{https://arxiv.org/abs/0902.0183}{{\ttfamily 0902.0183}}].

\bibitem{Beisert:2006ez}
N.~Beisert, B.~Eden and M.~Staudacher, \emph{Transcendentality and crossing},
  \href{https://doi.org/10.1088/1742-5468/2007/01/P01021}{\emph{J. Stat. Mech.}
  {\bfseries 0701} (2007) P01021}
  [\href{https://arxiv.org/abs/hep-th/0610251}{{\ttfamily hep-th/0610251}}].

\bibitem{Arutyunov:2006yd}
G.~Arutyunov, S.~Frolov and M.~Zamaklar, \emph{{The Zamolodchikov-Faddeev
  algebra for AdS(5) x S**5 superstring}},
  \href{https://doi.org/10.1088/1126-6708/2007/04/002}{\emph{JHEP} {\bfseries
  04} (2007) 002} [\href{https://arxiv.org/abs/hep-th/0612229}{{\ttfamily
  hep-th/0612229}}].

\bibitem{Arutyunov:2009kf}
G.~Arutyunov and S.~Frolov, \emph{The dressing factor and crossing equations},
  \href{https://doi.org/10.1088/1751-8113/42/42/425401}{\emph{J. Phys.}
  {\bfseries A42} (2009) 425401}
  [\href{https://arxiv.org/abs/0904.4575}{{\ttfamily 0904.4575}}].

\bibitem{Drummond:2006rz}
J.~M. Drummond, J.~Henn, V.~A. Smirnov and E.~Sokatchev, \emph{{Magic
  identities for conformal four-point integrals}},
  \href{https://doi.org/10.1088/1126-6708/2007/01/064}{\emph{JHEP} {\bfseries
  01} (2007) 064} [\href{https://arxiv.org/abs/hep-th/0607160}{{\ttfamily
  hep-th/0607160}}].

\bibitem{DHoker:1999jke}
E.~D'Hoker, D.~Z. Freedman, S.~D. Mathur, A.~Matusis and L.~Rastelli,
  \emph{{Extremal correlators in the AdS / CFT correspondence}},
  \href{https://arxiv.org/abs/hep-th/9908160}{{\ttfamily hep-th/9908160}}.

\bibitem{Bianchi:1999ie}
M.~Bianchi and S.~Kovacs, \emph{{Nonrenormalization of extremal correlators in
  N=4 SYM theory}},
  \href{https://doi.org/10.1016/S0370-2693(99)01211-3}{\emph{Phys. Lett.}
  {\bfseries B468} (1999) 102}
  [\href{https://arxiv.org/abs/hep-th/9910016}{{\ttfamily hep-th/9910016}}].

\bibitem{Eden:1999kw}
B.~Eden, P.~S. Howe, C.~Schubert, E.~Sokatchev and P.~C. West, \emph{{Extremal
  correlators in four-dimensional SCFT}},
  \href{https://doi.org/10.1016/S0370-2693(99)01442-2}{\emph{Phys. Lett.}
  {\bfseries B472} (2000) 323}
  [\href{https://arxiv.org/abs/hep-th/9910150}{{\ttfamily hep-th/9910150}}].

\bibitem{Chicherin:2015edu}
D.~Chicherin, J.~Drummond, P.~Heslop and E.~Sokatchev, \emph{{All three-loop
  four-point correlators of half-BPS operators in planar $ \mathcal{N} $ = 4
  SYM}}, \href{https://doi.org/10.1007/JHEP08(2016)053}{\emph{JHEP} {\bfseries
  08} (2016) 053} [\href{https://arxiv.org/abs/1512.02926}{{\ttfamily
  1512.02926}}].

\bibitem{Sfondrini:2014via}
A.~Sfondrini, \emph{Towards integrability for {$\AdS{3}/\CFT_{2}$}},
  \href{https://doi.org/10.1088/1751-8113/48/2/023001}{\emph{J. Phys.}
  {\bfseries A48} (2015) 023001}
  [\href{https://arxiv.org/abs/1406.2971}{{\ttfamily 1406.2971}}].

\bibitem{Baggio:2018gct}
M.~Baggio and A.~Sfondrini, \emph{{Strings on NS-NS Backgrounds as Integrable
  Deformations}},  \href{https://arxiv.org/abs/1804.01998}{{\ttfamily
  1804.01998}}.

\bibitem{Dei:2018mfl}
A.~Dei and A.~Sfondrini, \emph{{Integrable spin chain for stringy
  Wess-Zumino-Witten models}},
  \href{https://arxiv.org/abs/1806.00422}{{\ttfamily 1806.00422}}.

\end{thebibliography}

\makeatletter \@ifundefined{Sphere}{\newcommand{\Sphere}{\text{S}}}{}
  \@ifundefined{AdS}{\newcommand{\AdS}{\text{AdS}}}{}
  \@ifundefined{CFT}{\newcommand{\CFT}{\text{CFT}}}{}
  \@ifundefined{CP}{\newcommand{\CP}{\text{CP}}}{}
  \@ifundefined{Torus}{\newcommand{\Torus}{\text{T}}}{}
  \@ifundefined{superN}{\newcommand{\superN}{\mathcal{N}}}{}
  \@ifundefined{grpOSp}{\newcommand{\grpOSp}{\mathrm{OSp}}}{}
  \@ifundefined{grpPSU}{\newcommand{\grpPSU}{\mathrm{PSU}}}{}
  \@ifundefined{grpSU}{\newcommand{\grpSU}{\mathrm{SU}}}{}
  \@ifundefined{grpU}{\newcommand{\grpU}{\mathrm{U}}}{}
  \@ifundefined{grpD}{\newcommand{\grpD}{\mathrm{D}}}{}
  \@ifundefined{grpSL}{\newcommand{\grpSL}{\mathrm{SL}}}{}
  \@ifundefined{grpSp}{\newcommand{\grpSp}{\mathrm{Sp}}}{}
  \@ifundefined{grpUSp}{\newcommand{\grpUSp}{\mathrm{USp}}}{}
  \@ifundefined{grpSO}{\newcommand{\grpSO}{\mathrm{SO}}}{}
  \@ifundefined{grpO}{\newcommand{\grpO}{\mathrm{O}}}{}
  \@ifundefined{algOSp}{\newcommand{\algOSp}{\mathfrak{osp}}}{}
  \@ifundefined{algPSU}{\newcommand{\algPSU}{\mathfrak{psu}}}{}
  \@ifundefined{algSU}{\newcommand{\algSU}{\mathfrak{su}}}{}
  \@ifundefined{algSp}{\newcommand{\algSp}{\mathfrak{sp}}}{}
  \@ifundefined{algSL}{\newcommand{\algSL}{\mathfrak{sl}}}{}
  \@ifundefined{algGL}{\newcommand{\algGL}{\mathfrak{gl}}}{}
  \@ifundefined{algU}{\newcommand{\algU}{\mathfrak{u}}}{}
  \@ifundefined{algSO}{\newcommand{\algSO}{\mathfrak{so}}}{}
  \@ifundefined{algO}{\newcommand{\algO}{\mathfrak{o}}}{}
  \@ifundefined{Integers}{\newcommand{\Integers}{\mathbb{Z}}}{}
  \@ifundefined{Reals}{\newcommand{\Reals}{\mathbb{R}}}{} \makeatother

\providecommand{\href}[2]{#2}\begingroup\raggedright\endgroup

\end{document}